\documentclass[iop,linenumbers]{emulateapj}
\usepackage{color} 
\usepackage{ulem}
\usepackage{amsmath}
\usepackage{enumerate}
\usepackage{apjfonts}
\usepackage{subfigure}
\usepackage{float}
\usepackage{multirow}
\usepackage[colorlinks=true,citecolor=blue,breaklinks=true]{hyperref}
\usepackage[toc,page]{appendix}
\usepackage{soul}
\setstcolor{red}

\newcommand\ergcms{erg~cm$^{-2}$~s$^{-1}$}

\begin{document}

\title{Peculiar disk behaviors of the black hole candidate MAXI J1348-630 in the hard state observed by \textit{Insight}-HXMT and \textit{Swift}}

\author{W. Zhang\altaffilmark{1,2}, L. Tao\altaffilmark{1}, R. Soria\altaffilmark{3,4}, J. L. Qu\altaffilmark{1}, S. N. Zhang\altaffilmark{1,2}, S. S. Weng\altaffilmark{5}, L. zhang\altaffilmark{6}, Y. N. Wang\altaffilmark{6}, Y. Huang\altaffilmark{1}, R. C. Ma\altaffilmark{1,2}, S. Zhang\altaffilmark{1}, M. Y. Ge\altaffilmark{1}, L. M. Song\altaffilmark{1}, X. Ma\altaffilmark{1}, Q. C. Bu\altaffilmark{1}, C. Cai\altaffilmark{1,2}, X. L. Cao\altaffilmark{1}, Z. Chang\altaffilmark{1}, L. Chen\altaffilmark{1,7}, T. X. Chen\altaffilmark{1}, Y. B. Chen\altaffilmark{8}, Y. Chen\altaffilmark{1}, Y. P. Chen\altaffilmark{1}, W. W. Cui\altaffilmark{1}, Y. Y. Du\altaffilmark{1}, G. H. Gao\altaffilmark{1,2}, H. Gao\altaffilmark{1,2}, Y. D. Gu\altaffilmark{1}, J. Guan\altaffilmark{1}, C. C. Guo\altaffilmark{1,2}, D. W. Han\altaffilmark{1}, J. Huo\altaffilmark{1}, S. M. Jia\altaffilmark{1}, W. C. Jiang\altaffilmark{1}, J. Jin\altaffilmark{1}, L. D. Kong\altaffilmark{1,2}, B. Li\altaffilmark{1}, C. K. Li\altaffilmark{1}, G. Li\altaffilmark{1}, T. P. Li\altaffilmark{1,2,8}, W. Li\altaffilmark{1}, X. Li\altaffilmark{1}, X. B. Li\altaffilmark{1}, X. F. Li\altaffilmark{1}, Z. W. Li\altaffilmark{1}, X. H. Liang\altaffilmark{1}, J. Y. Liao\altaffilmark{1}, B. S. Liu\altaffilmark{1}, C. Z. Liu\altaffilmark{1}, H. X. Liu\altaffilmark{1,2}, H. W. Liu\altaffilmark{1}, X. J. Liu\altaffilmark{1}, F. J. Lu\altaffilmark{1}, X. F. Lu\altaffilmark{1}, Q. Luo\altaffilmark{1,2}, T. Luo\altaffilmark{1}, B. Meng\altaffilmark{1}, Y. Nang\altaffilmark{1,2}, J. Y. Nie\altaffilmark{1}, G. Ou\altaffilmark{1}, X. Q. Ren\altaffilmark{1,2}, N. Sai\altaffilmark{1,2}, X. Y. Song\altaffilmark{1}, L. Sun\altaffilmark{1}, Y. Tan\altaffilmark{1}, Y. L. Tuo\altaffilmark{1,2}, C. Wang\altaffilmark{2,9}, L. J. Wang\altaffilmark{1}, P. J. Wang\altaffilmark{1,2}, W. S. Wang\altaffilmark{1}, Y. S. Wang\altaffilmark{1}, X. Y. Wen\altaffilmark{1}, B. Y. Wu\altaffilmark{1,2}, B. B. Wu\altaffilmark{1}, M. Wu\altaffilmark{1}, G. C. Xiao\altaffilmark{2,10}, S. Xiao\altaffilmark{1,2}, S. L. Xiong\altaffilmark{1}, Y. P. Chen\altaffilmark{1,2}, R. J. Yang\altaffilmark{11}, S. Yang\altaffilmark{1}, Y. J. Yang\altaffilmark{1}, Y. R. Yang\altaffilmark{1}, Q. B. Yi\altaffilmark{1,12}, Q. Q. Yin\altaffilmark{1}, Y. Yuan\altaffilmark{1}, F. Zhang\altaffilmark{1}, H. M. Zhang\altaffilmark{1}, P. Zhang\altaffilmark{1}, W. C. Zhang\altaffilmark{1}, Y. F. Zhang\altaffilmark{1}, Y. H. Zhang\altaffilmark{1,2}, H. S. Zhao\altaffilmark{1}, X. F. Zhao\altaffilmark{1,2}, S. J. Zheng\altaffilmark{1}, Y. G. Zheng\altaffilmark{1,11}, D. K. Zhou\altaffilmark{1,2}}

\altaffiltext{1}{Key Laboratory of Particle Astrophysics, Institute of High Energy Physics, Chinese Academy of Sciences, Beijing 100049, China; \email{taolian@ihep.ac.cn; qujl@ihep.ac.cn}}
\altaffiltext{2}{University of Chinese Academy of Sciences, Chinese Academy of Sciences, Beijing 100049, People's Republic of China}
\altaffiltext{3}{School of Astronomy and Space Sciences, University of Chinese Academy of Sciences, Beijing 100049, People's Republic of China}
\altaffiltext{4}{Sydney Institute for Astronomy, School of Physics A28, The University of Sydney, Sydney, NSW 2006, Australia}
\altaffiltext{5}{Department of Physics and Institute of Theoretical Physics, Nanjing Normal University, Nanjing 210023, China}
\altaffiltext{6}{Physics and Astronomy, University of Southampton, Southampton, Hampshire SO17 1BJ, UK}
\altaffiltext{7}{Department of Astronomy, Beijing Normal University, Beijing 100049, People's Republic of China}
\altaffiltext{8}{Department of Physics, Tsinghua University, Beijing 100049, People's Republic of China}
\altaffiltext{9}{Key Laboratory of Space Astronomy and Technology, National Astronomical Observatories, Chinese Academy of Sciences, Beijing 100012, China}
\altaffiltext{10}{Key Laboratory of Dark Matter and Space Astronomy, Purple Mountain Observatory, Chinese Academy of Sciences, 210023 Nanjing, Jiangsu, China}
\altaffiltext{11}{School of Physics and Optoelectronics, Xiangtan University, Yuhu District, Xiangtan, Hunan, 411105, China}
\altaffiltext{12}{College of physics Sciences \& Technology, Hebei University,
No. 180 Wusi Dong Road, Lian Chi District, Baoding City, Hebei Province, 071002 China}
\shorttitle{MAXI J1348-630}
\shortauthors{Zhang et al.}

\begin{abstract}

We present a spectral study of the black hole candidate MAXI J1348$-$630 during its 2019 outburst, based on monitoring observations with \textit{Insight}-HXMT and \textit{Swift}. Throughout the outburst, the spectra are well fitted with power-law plus disk-blackbody components. In the soft-intermediate and soft states, we observed the canonical relation $L \propto T_{\rm in}^4$ between disk luminosity $L$ and peak colour temperature $T_{\rm in}$, with a constant inner radius $R_{\rm in}$ (traditionally identified with the innermost stable circular orbit). At other stages of the outburst cycle, the behaviour is more unusual, inconsistent with the canonical outburst evolution of black hole transients. In particular, during the hard rise, the apparent inner radius is smaller than in the soft state (and increasing), and the peak colour temperature is higher (and decreasing). This anomalous behaviour is found even when we model the spectra with self-consistent Comptonization models, which take into account the up-scattering of photons from the disk component into the power-law component. To explain both those anomalous trends at the same time, we suggest that the hardening factor for the inner disk emission was larger than the canonical value of $\approx$\,1.7 at the beginning of the outburst. A more physical trend of radii and temperature evolution requires a hardening factor evolving from $\approx$\,3.5 at the beginning of the hard state to $\approx$\,1.7 in the hard intermediate state. This could be evidence that the inner disk was in the process of condensing from the hot, optically thin medium and had not yet reached a sufficiently high optical depth for its emission spectrum to be described by the standard optically-thick disk solution.

\end{abstract}


\section{Introduction}
\label{sec:intro}
A black hole binary (BHB) consists of a black hole (BH) and a companion star. Depending on the mass of their donor stars, BHBs are classified as high mass X-ray binaries (HMXBs; ${\rm M_{2}}\ge10\,M_{\odot}$) or low mass X-ray binaries (LMXBs; ${\rm M_{2}}\le1\,M_{\odot}$). Matter is transferred from the companion star to the compact object mainly through stellar winds in HMXBs, and Roche-lobe overflow in LMXBs \citep{2006csxs.book..215C}, with the formation of a geometrically thin, optically thick accretion disk \citep{1973A&A....24..337S}.

Roche-lobe fed BHBs tend to be transient, alternating between outburst and quiescence. In the ``canonical'' interpretation, outbursts are due to disk instabilities \citep{Lasota2001}. Outbursts are typically divided into several spectral states \citep[e.g.,][]{2004MNRAS.355.1105F, 2006ARA&A..44...49R, 2009MNRAS.400.1603M}. At the beginning, the source rises in the hard state (HS), dominated by a hard power-law component in the X-ray band, and with a flat-spectrum radio core attributed to a steady compact jet. The power-law component extends to $>$100\,keV and originates from inverse Compton scattering in a hot corona or the base of the jet. The accretion disk is usually thought to be truncated and does not dominate the X-ray emission. In an X-ray hardness-intensity diagram (HID), the source at this point is rising along the right branch. Next, a transient BHB generally switches to the intermediate state (IMS). The soft thermal emission from the disk component gradually becomes significant, as the disk extends inward; the source moves along the top horizontal branch in the HID. Normally, the IMS can be further divided into the hard intermediate state (HIMS) and soft intermediate state (SIMS). The spectra of the SIMS are slightly softer than the HIMS, and the so-called type-B QPO would appear in the SIMS. When the jet turns off and the spectra are softer and dominated by the disk emission, the transient enters the high/soft (or thermal dominant) state (SS, top left section of the HID). In this state, the type-B QPO disappears and the inner disk is thought to reach the innermost stable circular orbit (ISCO). After several weeks or months, transient BHBs decline through an IMS (lower horizontal branch of the HID) and finally back to the low/hard state with a restarted jet, completing a counterclockwise q-shaped track in the HID \citep{2005A&A...440..207B, 2016ASSL..440...61B,2006ARA&A..44...49R}.

This simplified outburst outline may suggest that the physical evolution of disk, jet and corona are all connected, and produce the common behaviour in all BHBs. However, this may not always be the case. In this paper, we illustrate and discuss one example of X-ray outburst in a recently discovered BH candidate, MAXI J1348$-$630, in which the evolution of the fitted disk parameters is different from the canonical scenario. This suggests that there are still important aspects of the outburst cycle that differ from source to source and require more accurate modelling.

MAXI J1348$-$630 was discovered with the \textit{Monitor of All-sky X-ray Image} \citep[\textit{MAXI};][]{2009PASJ...61..999M} Gas Slit Camera (GSC) on 2019 January 26 \citep{2019ATel12425....1Y}. The initial discovery triggered a flurry of follow-up observations: with the {\it Hard X-ray Modulation Telescope} (\textit{Insight}-HXMT) \citep{Chen2019}, the {\it Neil Gehrels Swift Observatory} (\textit{Swift}) \citep{Kennea2019}, \textit{INTEGRAL} \citep{Lepingwell2019} and the {\it Neutron star Interior Composition Explorer} ({\it NICER}) \citep{2019ATel12447....1S} in the X-ray bands; with the 50-cm T31 telescope \citep{Denisenko2019}, the Las Cumbres Observatory 2-m and 1-m robotic telescopes \citep{2019ATel12439....1R} and the Southern African Large Telescope \citep{Charles2019} in the optical bands; with the Australia Telescope Compact Array \citep{2019ATel12456....1R}, MeerKAT \citep{Carotenuto2019} and the Murchison Widefield Array \citep{Chauhan2019} in the radio bands.
A distance of $\sim$2.2 kpc was estimated from the HI absorption \citep{Chauhan2020}, or of $\sim$3.39 kpc from the dust scattering ring around the source \citep{Lamer2020}.

The optical counterpart brightened by at least $\sim$2.8 mag at the beginning of the outburst \citep{Denisenko2019,2019ATel12439....1R}, as expected for an LMXB. There is no dynamical mass measurement for the compact object. Its identification as a BH candidate is supported by its radio detection \citep{2019ATel12456....1R, Carotenuto2019}, the strong broadband noise and ``power color'' in its power spectrum \citep{2019ATel12447....1S}, its optical/X-ray \citep{2019ATel12439....1R} and radio/X-ray \citep{2019ATel12456....1R} correlations. The main outburst lasted for about four months, with a peak flux of $\sim$4 Crab in the 2--20\,keV band \citep{Tominaga2020}, followed by several re-brightenings  \citep[e.g.,][]{2019ATel12829....1R,Negoro2019,AlYazeedi2019,Pirbhoy2020,Zhangl2020a}. A compact, steady radio jet was detected in the initial HS \citep{2019ATel12456....1R, Carotenuto2019}, and bright radio flaring marked the hard-to-soft state transition \citep{Carotenuto2019, Chauhan2019}. Low-frequency quasi periodic oscillations (QPOs) were detected in the hard and intermediate states \citep[Huang et al., in prep]{Chen2019, 2019ATel12447....1S, Jana2019, Jana2020, Belloni2020, Zhangl2020b}. The inner disk radius and temperature in the SS, and the X-ray luminosity at the state transitions are consistent with a $\sim$10-$M_{\odot}$ stellar BH \citep{Tominaga2020,Jana2020}, with large uncertainties function of BH spin, viewing angle and source distance. The long-term monitoring suggested that the hard X-rays and the optical emission changed simultaneously, and both leaded the soft X-rays by about 8-12 days \citep{2021ApJ...915L..15W}.

In this paper, we present detailed results of our study of the outburst evolution with {\it Insight-HXMT}. The broad energy band and large effective area at high energy make {\it Insight-HXMT} an ideal instrument to study the spectral and timing evolution of X-ray transients, especially for bright sources, such as MAXI J1348$-$630, which may suffer from pile-up in other X-ray telescopes. We describe observations and data analysis in Section~\ref{sec:obs}, we model the spectral state evolution in Section~\ref{sec:res}, and discuss the results in Section~\ref{sec:dis}.

\section{Observations and data reduction}
\label{sec:obs}
\textit{Insight}-HXMT is the first Chinese X-ray observatory, launched on 2017 June 15 \citep{2020SciChina249502}. It carries three telescopes with high time resolutions and large effective areas: the Low Energy (LE) telescope with an effective area of 384 cm$^{2}$ at 1--15\,keV \citep{2020SciChina249505}; the Medium Energy (ME) telescope with 952 cm$^{2}$ at 5--30\,keV \citep{2020SciChina249504}; the High Energy (HE) telescope with 5100 cm$^{2}$ at 20--250\,keV \citep{2020SciChina249503}. Each telescope carries large and small field-of-view (FOV) detectors, in order to facilitate the background analyses. The small FOV detectors (LE: $1.6^{\circ} \times 6^{\circ}$; ME: $1^{\circ} \times 4^{\circ}$; HE: $1.1^{\circ} \times 5.7^{\circ}$) have low probability of source contamination, thus are applicable to observe point sources.

\textit{Insight}-HXMT started to observe MAXI J1348$-$630 on 2019 January 27 (MJD 58510), one day after its discovery, and monitored it until the end of the main outburst on May 30. We obtained 59 observations, each divided into sub-exposures (a total of 171 individual sub-exposures). The exposure time of each observation varied between 10--150 ks, with a total exposure time of 1500 ks.  

We extracted and analyzed the data with the \textit{Insight}-HXMT Data Analysis software\footnote{http://hxmten.ihep.ac.cn/software.jhtml} ({\sc HXMTDAS}) V2.04. We used only data from the small FOV detectors. We created good time intervals with the following selection criteria: (1) pointing offset angle $<$0.04$^{\circ}$; (2) elevation angle $>$10$^{\circ}$; (3) geomagnetic cutoff rigidity $>$8 GV; (4) at least 300 s away from the crossing of the South Atlantic Anomaly. The LE, ME and HE background spectra are taken from the blind detectors because the spectral shape of the particle background is the same for both the blind and the small field-of-view detectors \citep{Liao2020a, Guo2020, Liao2020b}. We created background spectra with the {\it lebkgmap}, {\it mebkgmap} and {\it hebkgmap} tasks in {\sc HXMTDAS}, and response files with {\it lerspgen}, {\it merspgen} and {\it herspgen}.

The \textit{Swift} X-Ray Telescope (XRT) also monitored the main outburst, with 43 observations\footnote{The \textit{Swift} data were downloaded from https://heasarc.gsfc.nasa.gov/cgi-bin/W3Browse/w3browse.pl.}, each with a typical exposure time of $\sim$1 ks. We reprocessed the \textit{Swift}/XRT data with the {\sc xrtpipeline} software (calibration database version 20191017) and extracted spectra with {\it xselect} v2.4g. In most observations, the count rates are very high, above 100 ct s$^{-1}$. To mitigate that, we only used data taken in windowed timing mode; even in this mode, the pile-up effect cannot be ignored and the source core should be excluded from spectral extraction. Thus, we used annular source regions \citep{Tao2018} with an outer radius of 20 pixels and an inner radius adjusted to have a count rate below 100 ct s$^{-1}$. The background regions were chosen as annuli centered on the source, with an outer radius of 110 pixels and an inner radius of 90 pixels. Heavily absorbed sources (column densities of about $10^{22} \rm{cm^{-2}}$ and above) observed in windowed timing mode can show a spurious emission bump in the low energy spectrum, below 1 keV. To reduce this problem, we only included grade-0 events when extracting the spectrum\footnote{https://www.swift.ac.uk/analysis/xrt/digest\_cal.php.} and ignored energies below 0.6 keV. We generated ancillary response files with the {\it xrtmkarf} task in the {\sc ftools} package \citep{blackburn95}. 

We used the {\sc ftools} task {\it grppha} to rebin both sets of spectra (from {\it Insight}-HXMT and {\it Swift}) to a minimum of 20 counts per bin, and we fitted them in {\sc Xspec} \citep{Arnaud1996} with the $\chi^2$ statistics. We added a systematic uncertainty of 2\% to the {\it Insight}-HXMT spectra.

\section{Results}
\label{sec:res}
\subsection{Light Curve, Hardness Ratio and HIDs}

\begin{figure*}[tb]
    \centering
    \includegraphics[scale=0.8,trim=40 20 0 0]{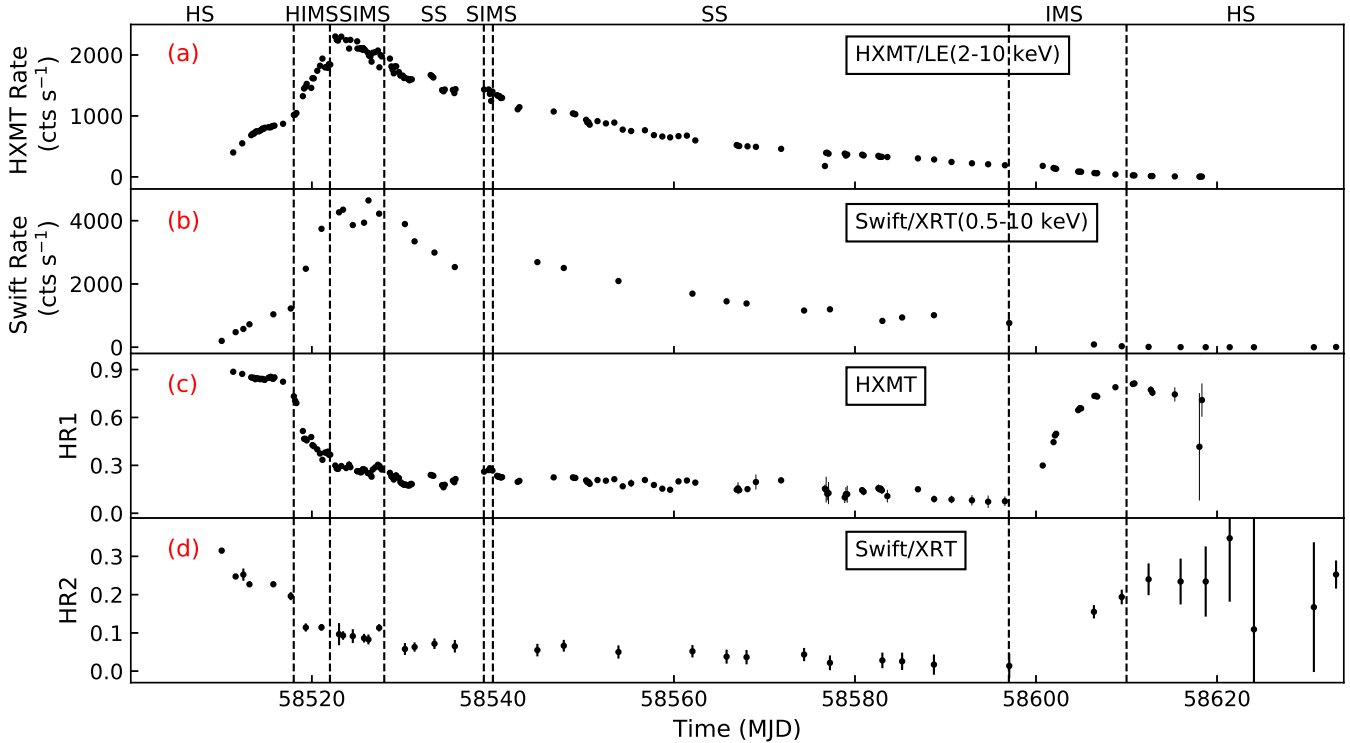}
    \caption{(a) \textit{Insight}-HXMT/LE 2--10\,keV light-curve. 
    (b) \textit{Swift}/XRT 0.5--10\,keV light-curve; we avoid pile-up problems by using an annular source extraction region, then correct the count rates with the ancillary response function. 
    (c) \textit{Insight}-HXMT/LE hardness ratio, defined as the ratio of (4--10\,keV)/(2--4\,keV) count rates. 
    (d) \textit{Swift}/XRT hardness ratio, defined as the ratio of (4--10\,keV)/(0.5--4\,keV) count rates. Vertical lines mark the onset of different states: hard state (HS), hard-intermediate state (HIMS), soft-intermediate state (SIMS), soft state (SS), and intermediate state (IMS) in the decline phase.}
    \label{fig:lc}
\end{figure*}

The \textit{Insight}-HXMT/LE 2--10\,keV light-curve shows (Figure~\ref{fig:lc}, panel a) an outburst peak at MJD 58522, when the LE count rates reached $\sim$2400 ct s$^{-1}$, followed by a gradual decline down to a count rates of 2 ct~s$^{-1}$ on its last observation of the source (MJD 58617). The evolution is consistent with the \textit{Swift}/XRT light-curve (Figure~\ref{fig:lc}, panel b), with the added bonus that the \textit{Insight}-HXMT/LE observations have higher cadence and no pile-up issues even at outburst peak.

\begin{figure}
        \includegraphics[scale=0.6,trim=20 10 10 0]{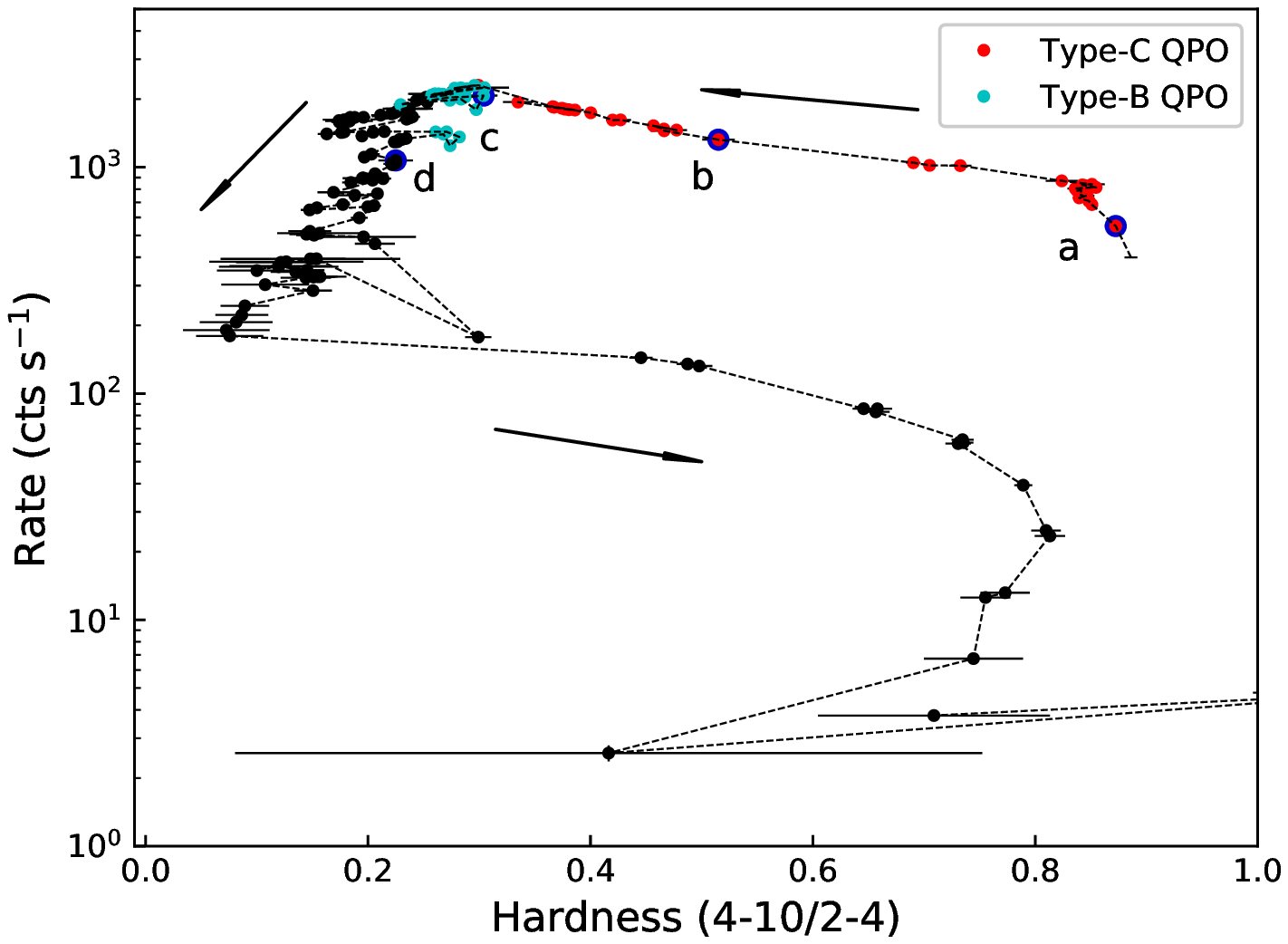}\\[-7pt]
        \includegraphics[scale=0.6,trim=20 10 10 0]{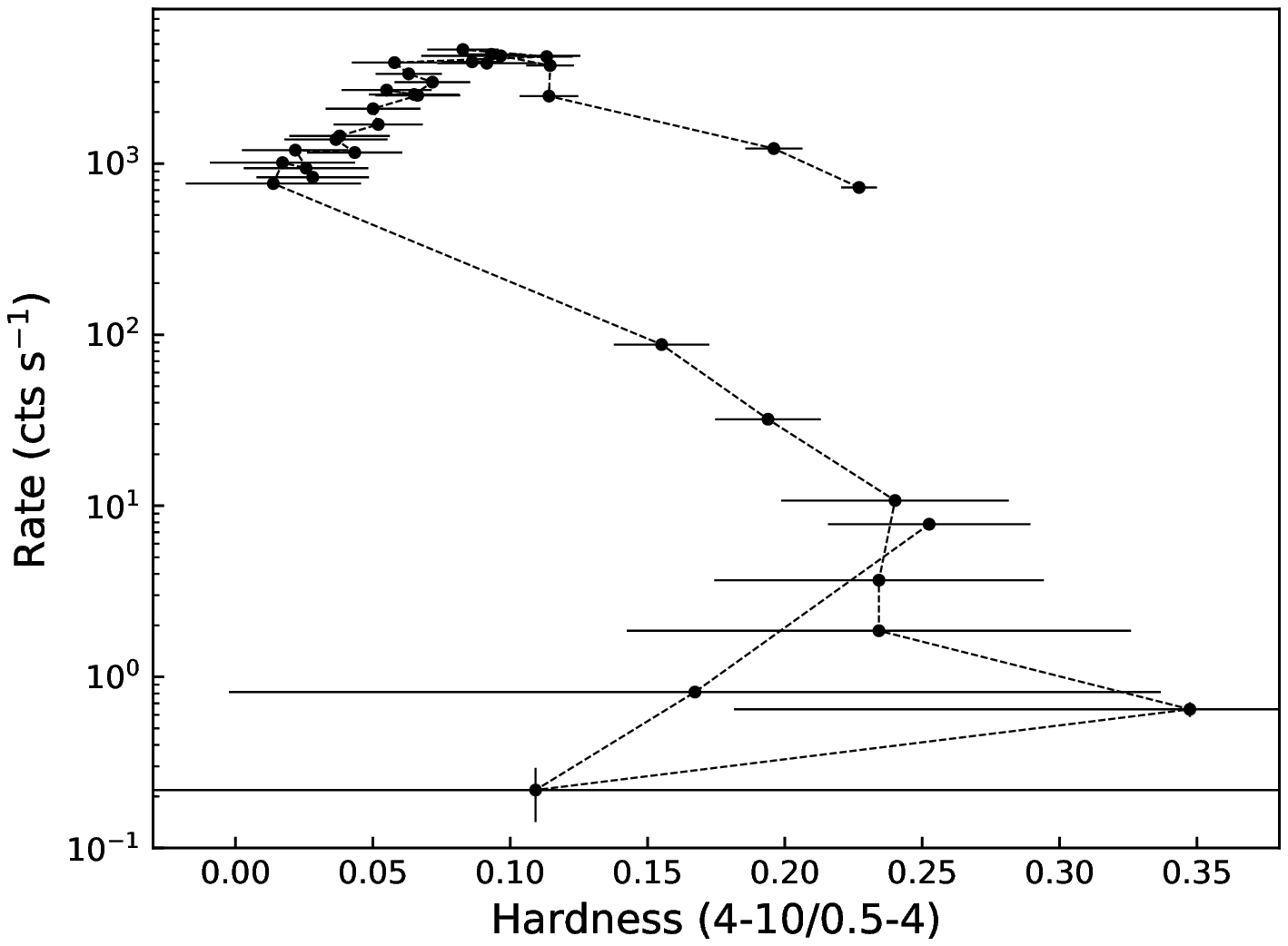}\\[-7pt]
        \includegraphics[scale=0.6,trim=20 10 10 0]{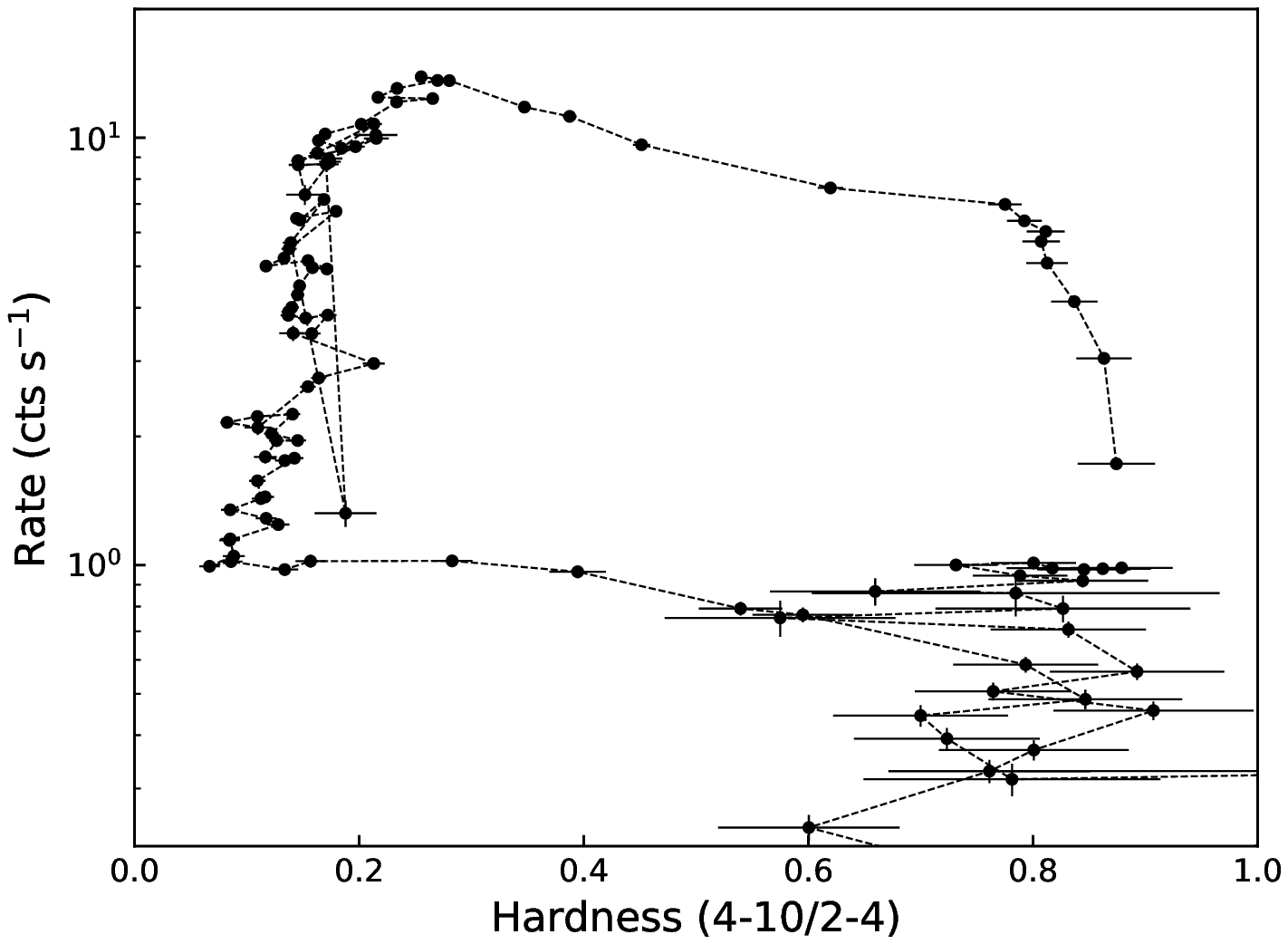}
    \caption{Top panel: HID based on the data from \textit{Insight}-HXMT/LE. Red points mark the observations with Type-C QPOs, cyan points those with type-B QPOs, and four blue circles marked with 'a', 'b', 'c' and 'd' corresponding to the four observations in Figure 3. Middle panel: HID from \textit{Swift}/XRT. Bottom panel: HID from \textit{MAXI}/GSC.}
    \label{fig:hids}
\end{figure}

We define a hardness ratio HR1 from the ratio of the 4--10\,keV over the 2--4\,keV count rates in \textit{Insight}-HXMT/LE, and a hardness ratio HR2 for the 4--10\,keV over 0.5--4\,keV rates in \textit{Swift}/XRT. Both HR1 and HR2 clearly show (Figure~\ref{fig:lc}, panels c,d) evolution and a sequence of transitions, corresponding to various outburst stages. The difference between HIMS and SIMS is defined by the detection of a Type-B QPO in the latter, from MJD 58522 to MJD 58528 and again very briefly from MJD 58539 to MJD 58540 \citep[Huang et al., in prep.]{Belloni2020,Zhangl2020b}. \textit{Insight}-HXMT/LE's data show very clearly that the transition from HIMS to SIMS corresponds to the peak count rates. From MJD 58528 to MJD 58596 is the SS, during which the flux decays exponentially (Figure~\ref{fig:lc}, panel a). The decay timescale is 32 days. The reverse transition from the SS to the IMS and then HS starts on MJD 58596. Combining the intensities and hardness ratios, we obtain the typical q-shaped evolutionary track expected for a BHB outburst (Figure~\ref{fig:hids}), with consistent results between \textit{Insight}-HXMT/LE and \textit{Swift}/XRT. For comparison, we also plot the HID from \textit{MAXI}/GSC\footnote{The \textit{MAXI} data are obtained from http://maxi.riken.jp/mxondem/.}; see \citet{Tominaga2020} for the \textit{MAXI} results.

\subsection{Spectral Analysis}
\label{sec:spec}

\begin{figure*}[htb!]
    \subfigure
    {
    \centering
    \includegraphics[width=.5\textwidth,trim=20 20 10 0]{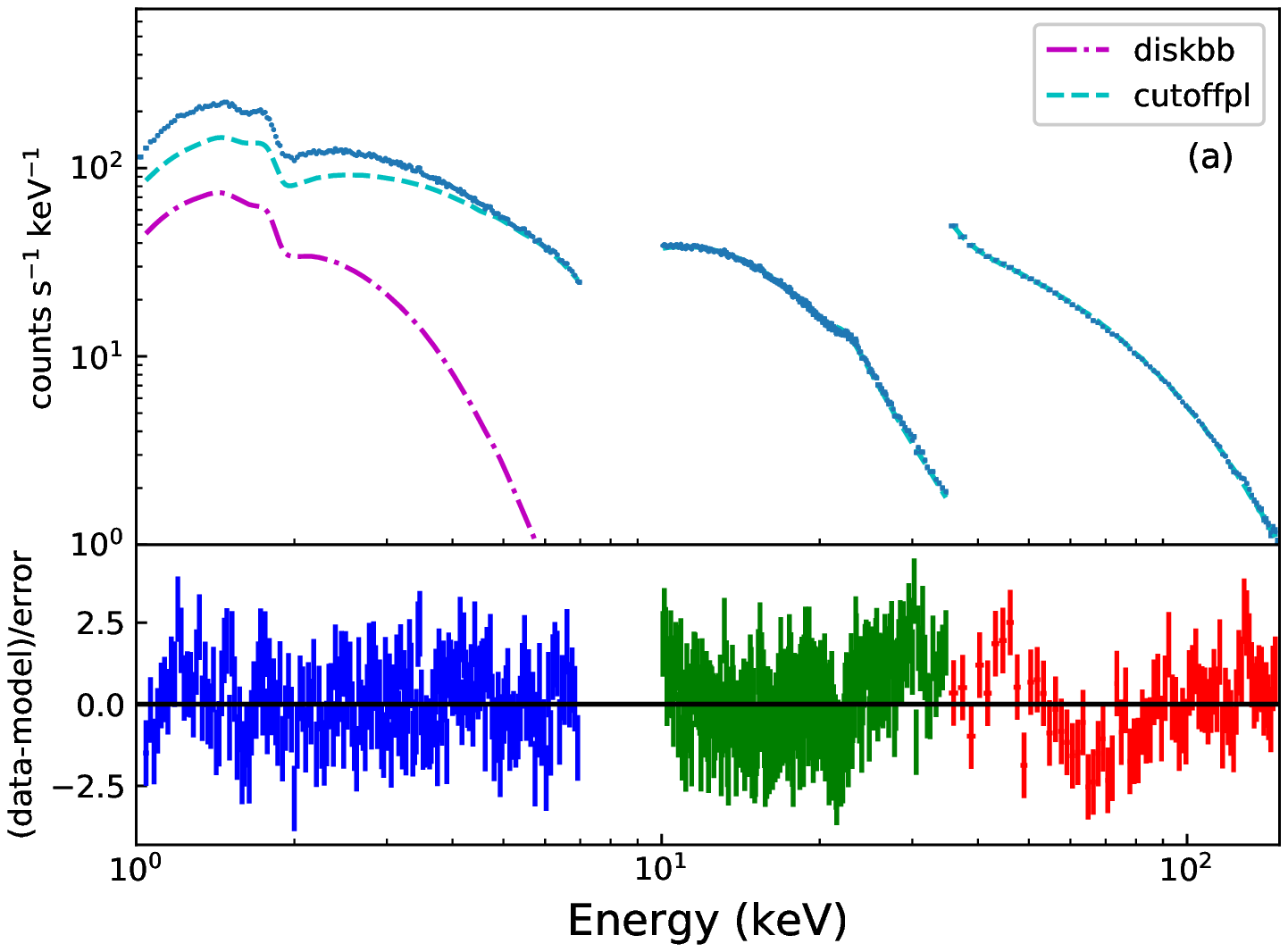}
    }
    \subfigure
    {
    \centering
    \includegraphics[width=.5\textwidth,trim=20 20 10 0]{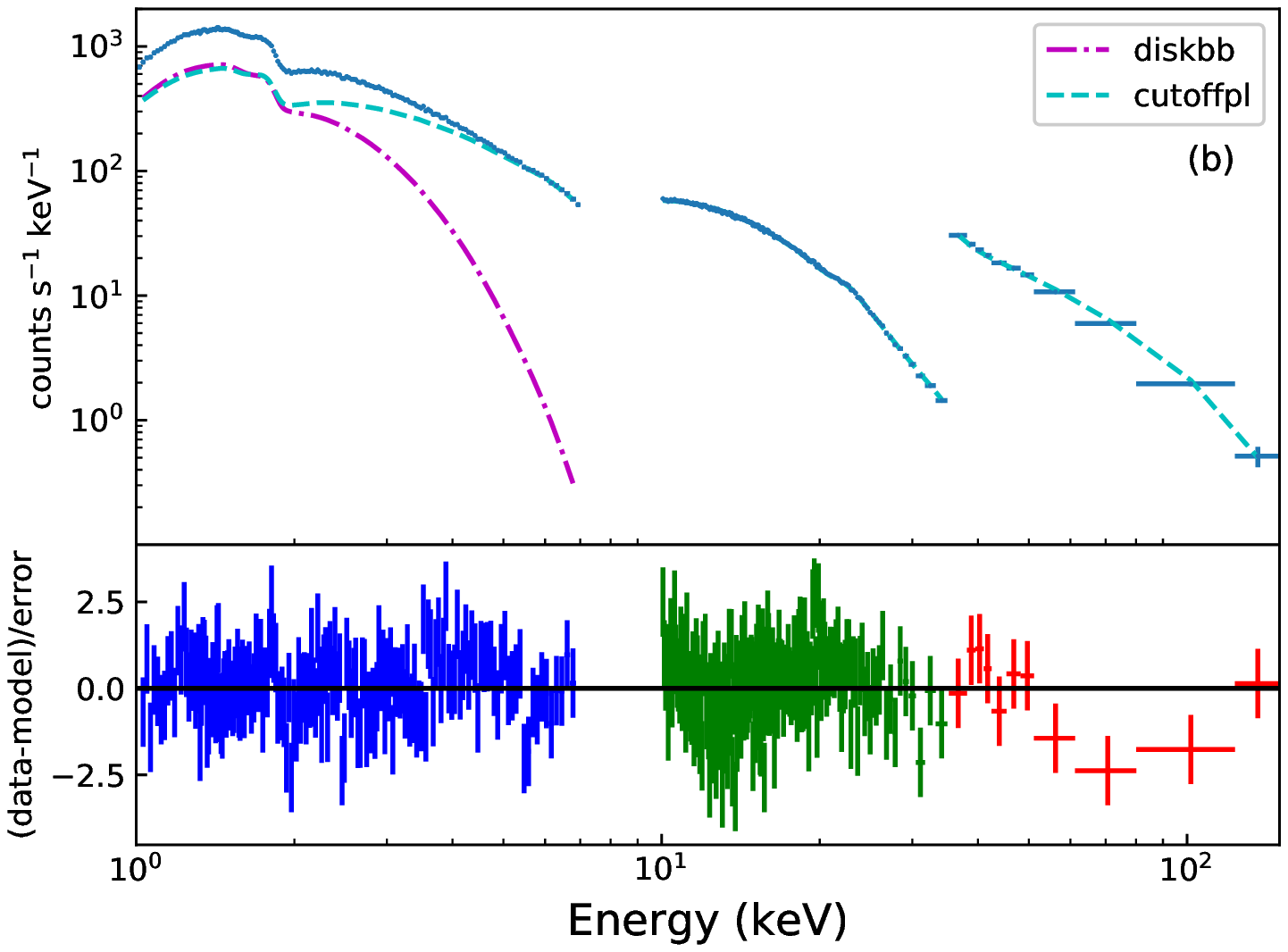}
    }
    \subfigure
    {
    \centering
    \includegraphics[width=.5\textwidth,trim=20 20 10 0]{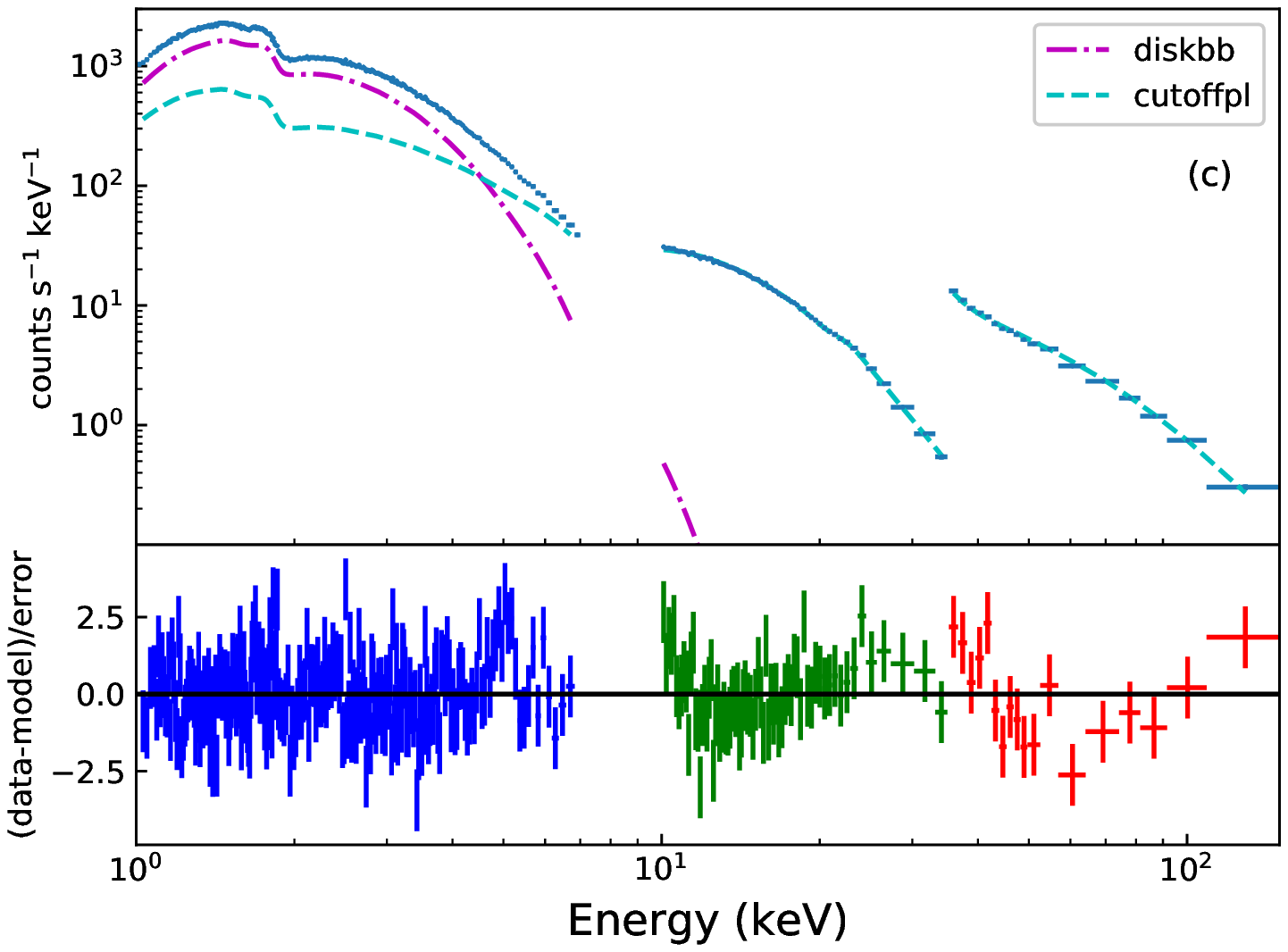}
    }
    \subfigure
    {
    \centering
    \includegraphics[width=.5\textwidth,trim=20 20 10 0]{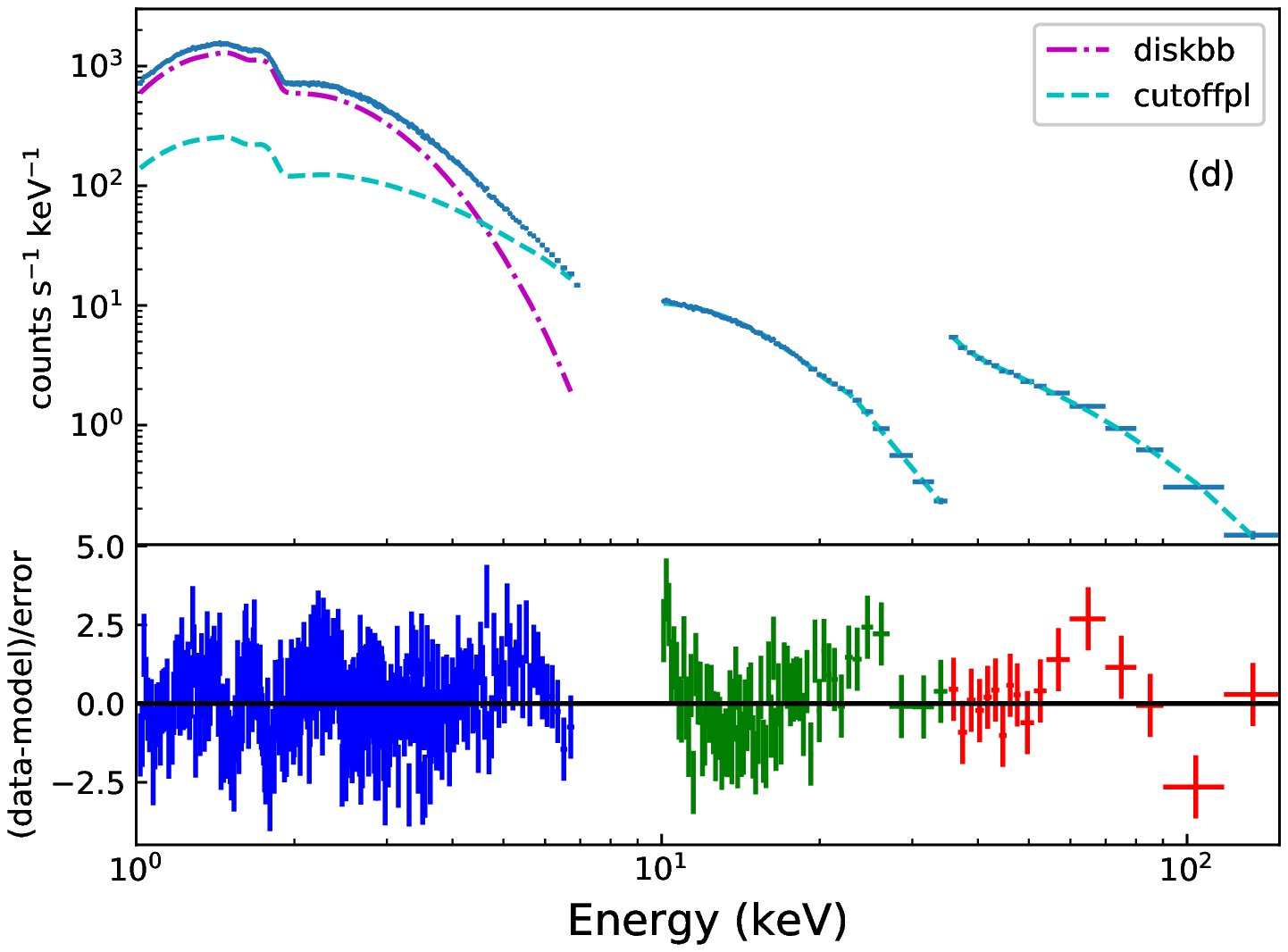}
    }
    \caption{Four representative spectra for different states of MAXI J1348-630, obtained from \textit{Insight}-HXMT observations 021400200201 (panel a, HS); 021400200701 (panel b, HIMS); 021400201701 (panel c, SIMS); 021400202801 (panel d, SS). The blue symbols for LE (1.0--7.0\,keV), the green symbols for ME (10.0--35.0\,keV), and the red symbols for HE (35.0--150.0\,keV). The {\tt diskbb} and {\tt cutoffpl} components are plotted in magenta dashed and cyan dash-dotted lines, respectively.}
    \label{fig:4spec}
\end{figure*}

\begin{figure*}[htb!]
    \centering
    \includegraphics[scale=0.85,trim=55 0 15 0]{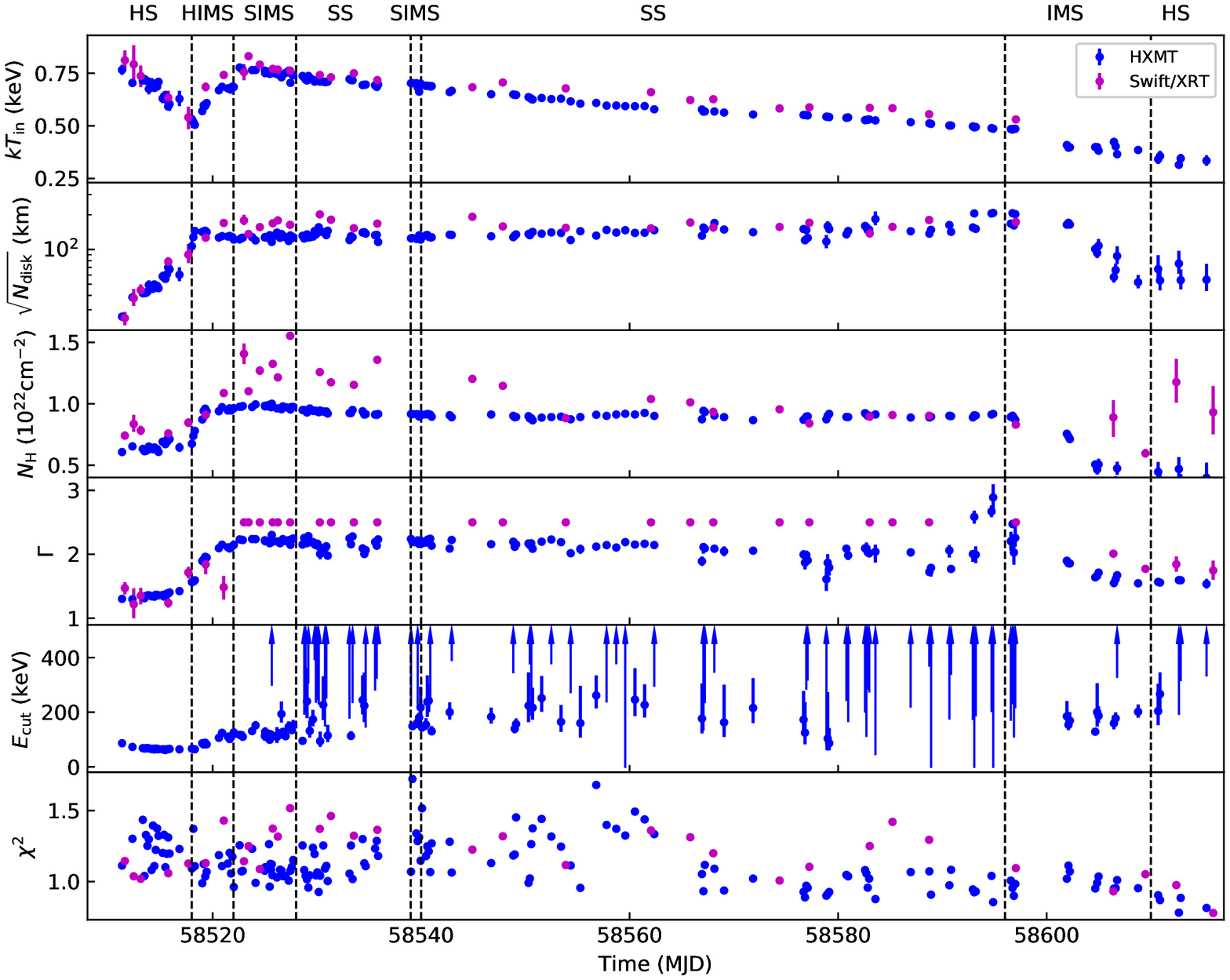}
    \caption{Evolution of \textit{Insight}-HXMT (blue points) and \textit{Swift} (magenta points) spectral parameters. From top to bottom: the inner disk color temperature ($T_{\rm in}$), the square root of the normalization of the {\tt diskbb} model ($N_{\rm disk}=(r_{\rm in}/D_{10})^{2}*{\rm cos}\theta$), the hydrogen absorption column density ($N_{\rm H}$), the power-law photon index ($\Gamma$) and the exponential roll-off energy ($E_{\rm cut}$).\\}
    \label{fig:spec}

    \centering
    \subfigure
    {
        \begin{minipage}{8cm}
        \centering
        \includegraphics[scale=0.56,trim=45 0 20 0]{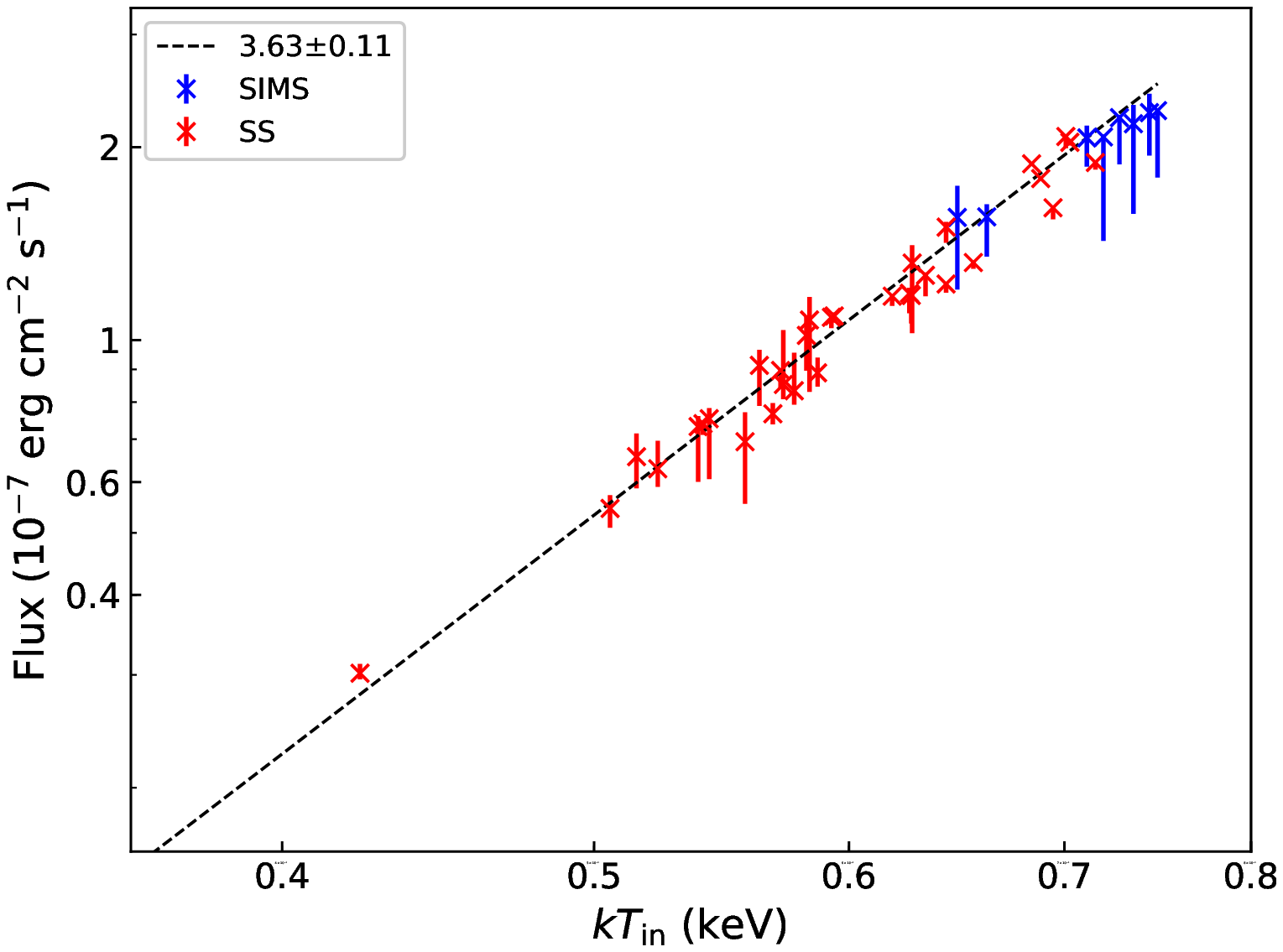}
        \end{minipage}
    }
    \subfigure
    {
        \begin{minipage}{8cm}
        \centering
        \includegraphics[scale=0.56,trim=15 0 20 0]{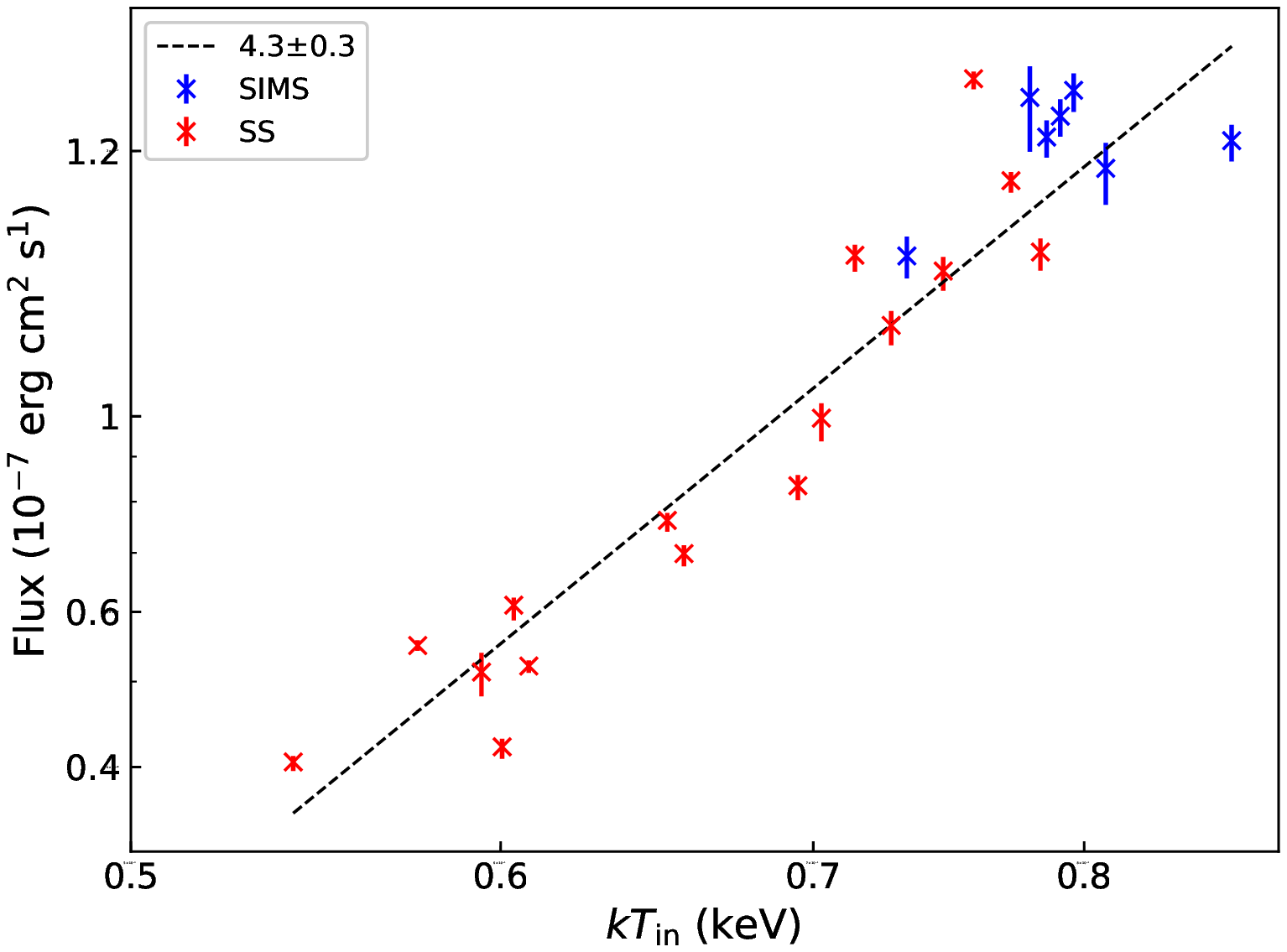}
        \end{minipage}
    }
    \caption{Disk flux versus inner disk temperature of the SIMS (blue symbols) and the SS (red symbols) for \textit{Insight}-HXMT (left panel) and \textit{Swift} (right panel). The flux is in units of \ergcms, and $kT_{\rm in}$ is in units of keV. The dashed lines represent the best fitting relation.\\}
    \label{fig:lxtin}
\end{figure*}

Not all observations provide enough signal-to-noise for detailed spectral modelling.
For \textit{Insight}-HXMT, we do not model the hard-state data from the fading stage (after MJD 58617). The observation on MJD 58510 (2019 January 27) is also not used, because its effective exposure time is too short (200 s). All \textit{Insight}-HXMT observations used for spectral analysis are listed in Table~\ref{tab:HXMTobs}. We also discard LE data in the 7--10\,keV band, because its systematic error is twice as large as in the 1--7\,keV band \citep{2020arXiv200306998L}. The energy bands used for LE, ME and HE fitting are 1--7\,keV, 10--35\,keV and 35--150\,keV, respectively. For \textit{Swift}/XRT, we use all available observations  (Table~\ref{tab:Swiftobs}), and fit the spectra in the 0.6--10\,keV band. All errors represent the 90\% confidence range for a single parameter, unless otherwise stated.

\subsubsection{{\tt diskbb+cutoffpl} model}
\label{sec:diskbb+po}
For a typical BHB, the continuum consists of a non-thermal component from the corona and a multicolor blackbody component from the disk. The power-law component dominates in the HS, the disk component in the SS. In the IMS, both components may give a significant contribution. Thus, we tried fitting the spectra with a single broad-band non-thermal component ({\tt cutoffpl} for \textit{Insight}-HXMT and {\tt powerlaw} for \textit{Swift}), a single disk blackbody ({\tt diskbb}) and a two-component model ({\tt diskbb+cutoffpl} for \textit{Insight}-HXMT and {\tt diskbb+powerlaw} for \textit{Swift}). 
For \textit{Insight}-HXMT, we included a free energy-independent multiplicative factor ({\tt constant}) to account for slightly different normalizations of LE, ME and HE.

The reason we used an exponential roll-off model for \textit{Insight}-HXMT but a simple power-law model for \textit{Swift} is that the cut-off energy is $>$ 60 keV and cannot be constrained in the 0.6--10\,keV XRT energy band. For the HS and HIMS, we used F-tests to determine whether the thermal component was significant. For the \textit{Swift} spectra in the SIMS and SS, we fixed the power-law photon index $\Gamma$ to the constant ``canonical'' value of 2.5 because the power-law tail in those states is very weak and its slope is unconstrained over the small XRT energy band.

Another possible source of X-ray emission that, in principle, we should be concerned about, for the LE (given its large field of view), is scattered emission from the dust ring discovered by \citet{Lamer2020}. Such structure is located at a distance of $\sim$2 kpc from us, it has an apparent angular radius on the sky of $\sim$40$^{\prime\prime}$ ($\sim$20 pc), and scatters X-ray photons into our line of sight with a time delay of $\sim$400 d from the original time of emission from MAXI J1348$-$630 \citep{Lamer2020}.
However, our observations cover only the first $\sim$100 days of the outburst, during which that particular scattering structure is not relevant. In principle, there may be other scattering rings already appearing within the first $\sim$100 days, caused by dust located at smaller angular distances from our line of sight to the source. Scaling from the results of \citet{Lamer2020}, and assuming those other putative dust structures are also located at $\sim$2 kpc, we estimate that a scattering ring visible after a time delay of $\sim$100 days would have a radius of $\sim$17$^{\prime}$ (a little  larger than the {\it Swift}/XRT field of view), or a radius of $\sim$12$^{\prime}$ ({\it Swift}/XRT field of view) for a time delay of $\sim$50 days. We searched for such rings of scattered emission in the {\it Swift} images but found no evidence. Thus, we can plausibly assume that the X-ray photons detected in our spectra are all from direct emission.

We modelled the absorption with {\tt tbabs}, with abundances from \citet{2000ApJ...542..914W}. Initially, we kept the column density $N_{\rm H}$ free at every epoch. Later, we repeated the modelling with a fixed value of $N_{\rm H}$ for the whole outburst (as discussed in Section 3.3).

Both disk and power-law components are needed for all observations, except for the last three \textit{Swift} observations near the end of the main outburst given the lower statistics. Four representative spectra for different state, obtained from \textit{Insight}-HXMT, are shown in Figure~\ref{fig:4spec}. The spectral evolution is shown in Figure~\ref{fig:spec}, and the spectral parameters are summarized in Tables~\ref{tab:HXMTobs} (\textit{Insight}-HXMT) and \ref{tab:Swiftobs} (\textit{Swift}). Most spectra can be well described by two continuum components, except for some \textit{Insight}-HXMT observations in the SS and SIMS with reduced $\chi^2$ larger than 1.5. Upon closer inspection, we find that the large values of $\chi^2$ come mainly from some narrow residuals in LE and the residuals in different observations have a similar structure, which may be due to some systematic uncertainties in calibration. As we focus on the evolution of the broad continuum components, the narrow residuals only have minor impact on the parameter determination.

The results from \textit{Insight}-HXMT and \textit{Swift} show a good consistency in the evolution for most parameters, especially for the disk parameters. However, we also note that $N_{\rm H}$ values in some \textit{Swift} observations are larger than those obtained from \textit{Insight}-HXMT, even after we adopt $N_{\rm H}$ derived from the 0.6--10\,keV band fits (see above). The photon index ($\Gamma$) is smaller in some \textit{Swift} spectra; this may be due to the different energy bands of the two satellites (1--150\,keV for \textit{Insight}-HXMT and 0.6--10\,keV for \textit{Swift}). We also test the fits by fixing \textit{Swift}-derived $\Gamma$ at the value of 2, and find that the other parameters change only slightly. Moreover, our results are consistent with previously published results: \citet{2019ATel12477....1B} analyzed the \textit{Swift} spectra from MJD 58517 and 58519, and obtained similar disk temperatures ($T_{\rm in}$) to ours. Our results also agree with $T_{\rm in}$ derived from the \textit{MAXI}/GSC spectra during the whole main outburst and justify their assumption about $\Gamma$ \citep{Tominaga2020}. The spectral evolution is also consistent with the \textit{NICER} results from \citet{Zhangl2020b} except for the disk normalization in the IMS at the end of the outburst. We also note that an extra {\tt gaussian} model is added in their spectral fitting in order to account for the extra residuals below 3\,keV after MJD 58603. The residuals may be due to the instrumental effects and the use of the {\tt gaussian} model may affect the determination of the disk normalization.

As show in Figure~\ref{fig:spec}, the parameters show a clear evolution between different states, which also justifies the classification of the spectral states. In the rising phase, the inner disk temperature ($kT_{\rm in}$) decreases from 0.75 to 0.5\,keV in the HS, and increases to 0.75\,keV in the HIMS. Then, it gradually decreases in the SIMS and SS, and suddenly decreases in the IMS and HS at the end of the outburst. The disk normalization ($N_{\rm disk}$) has an increasing trend in the HS and slightly decreases in the HIMS in the rising phase, then keeps nearly constant in the SIMS and SS. In the IMS and HS at the end of the outburst, $N_{\rm disk}$ gradually decreases. $\Gamma$ is small in the HS and gradually increases from 1.5 to 2.1 in the HIMS. $\Gamma$ then stays at 2.3 in the SIMS and SS, and decreases in the IMS and HS at the end of the outburst. $E_{\rm cut}$ ranges around 70\,keV in the HS, and then slightly increases. However, as the power-law component is weak in the SS and the source count rate is low during the fading stage, $E_{\rm cut}$ can not be well constrained since the SS. $N_{\rm{H}}$ also evolves in the outburst. $N_{\rm{H}}$ in the SS is larger than that in the HS, increasing from 0.7 to $1 \times \rm 10^{22}\ cm^{-2}$. We are aware that there is often a degeneracy between the fitted values of $N_{\rm{H}}$, $T_{\rm in}$ and $N_{\rm disk}$. We will examine and discuss the behaviour of those parameters under the assumption of a constant $N_{\rm{H}}$ in Section~\ref{sec:fixnh}.

The relative disk contribution to the unabsorbed flux increases from less than 10\% in the HS to above 60\% in the SS, and then decreases to below 10\% in the HS at the end of the outburst. The relations between the disk luminosity ($L_{\rm disk}$) and the inner disk color temperature ($T_{\rm in}$) during the SIMS and SS are shown in Figure~\ref{fig:lxtin}. $L_{\rm disk}$ is proportional to $T_{\rm in}^{3.63\pm 0.11}$ and $T_{\rm in}^{4.3\pm0.3}$ for \textit{Insight}-HXMT and \textit{Swift}, respectively, consistent with $L_{\rm{disk}}{\propto}T^{4}_{\rm{in}}$~expected from a standard thin disk extending to the ISCO \citep{1986ApJ...308..635M}. For other states, $L_{\rm disk}$ and $T_{\rm in}$ deviate significantly from the relationship of $L_{\rm{disk}}{\propto}T^{4}_{\rm{in}}$.

\subsubsection{{\tt diskbb+nthcomp+nthcomp} model}
\label{sec:2nth}
We find that, in the HS and the IMS, there is a bulge or a change of slope around 20--40\,keV in the spectra, when fitted with a single power-law component. This may be evidence of an additional Comptonization component. By analogy with the two Comptonizing regions assumed in the model of \cite{Garcia2021}, we also tried a double Comptonization model. First, we tried with two {\tt nthcomp} components (this Section), then  with one {\tt nthcomp} and one thermal comptonization component (Sections~\ref{sec:simplcut} and~\ref{sec:diskir}). In all our modelling, we fixed the seed photon temperature of {\tt nthcomp} at 0.1\,keV\footnote{We also tested the fits by linking the seed photon temperature of {\tt nthcomp} to $T_{\rm in}$ of {\tt diskbb} and find that the evolution trends do not change.} An additional {\tt diskbb} component was also included, because it is statistically significant (F-test null probability $<0.01$) both in the HS and the IMS. As shown in Figure~\ref{fig:Comptonspec} top panel, the {\tt diskbb} plus two {\tt nthcomp} model improved the spectral fits significantly, both in the 20--40\,keV region and in the shape of the down-up around 100\,keV. The two Comptonization component have different best-fitting values of $kT_{\rm e}$. The higher temperature is greater than 40\,keV, and may come from the region of the corona closer to the disk. The lower temperature is less than 15\,keV, and may come from the outer, cooler part of the corona. The detailed interpretation will be discussed in Guan et al. (in prep). The peak disk temperature and normalization are shown in the top panel of Figure~\ref{fig:additonal}: $N_{\rm disk}$ is consistent with that obtained with the simpler model (Section~\ref{sec:diskbb+po}), while $kT_{\rm in}$ is about 0.05\,keV lower, during the HS. The best-fitting values of the parameters and their uncertainties are shown in Table~\ref{tab:2nth}.

\subsubsection{{\tt simplcut*diskbb+nthcomp} model}
\label{sec:simplcut}
\begin{figure}
    \includegraphics[scale=0.57,trim=20 0 0 17]{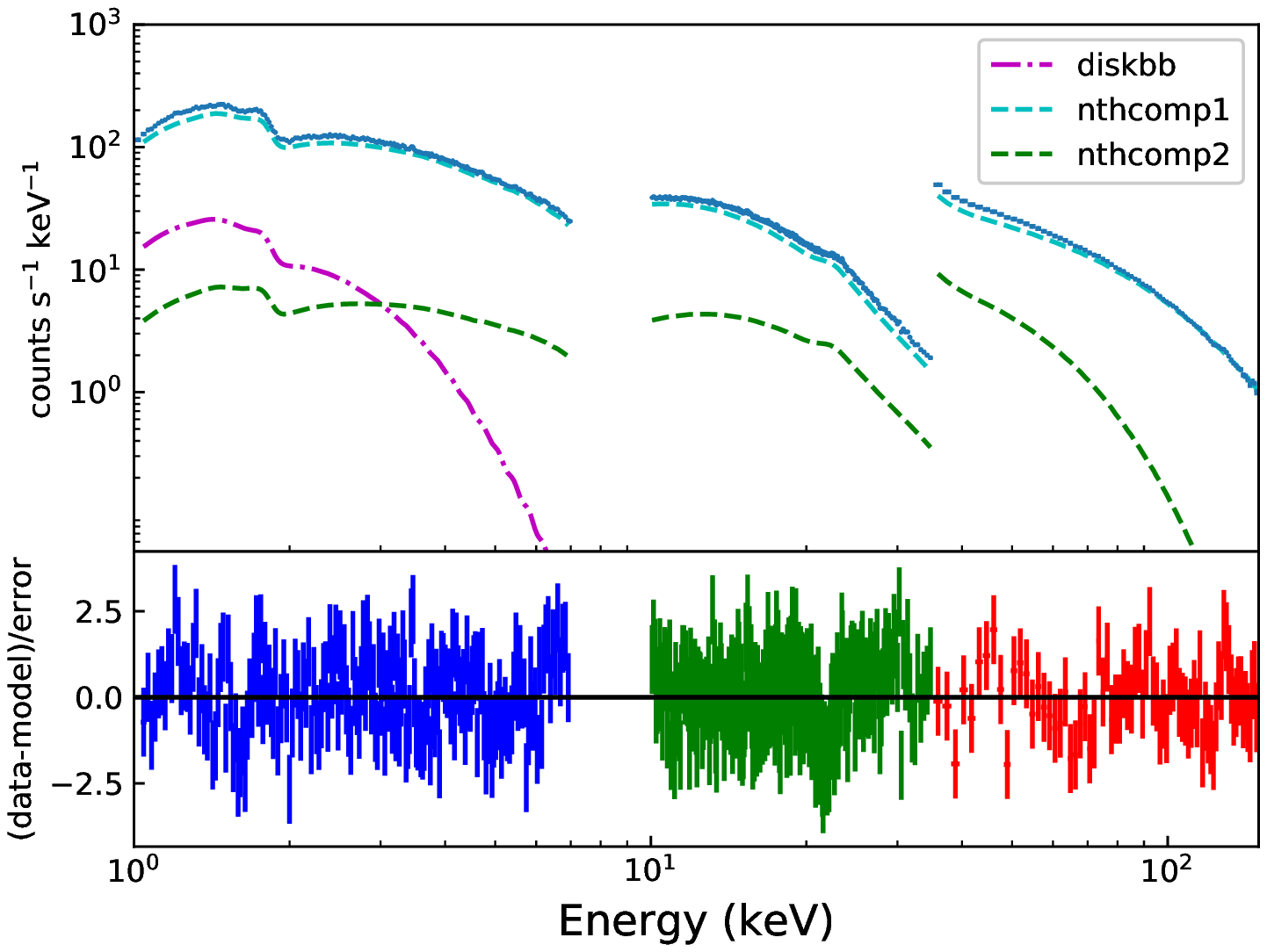}\\[-7pt]
    \includegraphics[scale=0.57,trim=20 10 0 10]{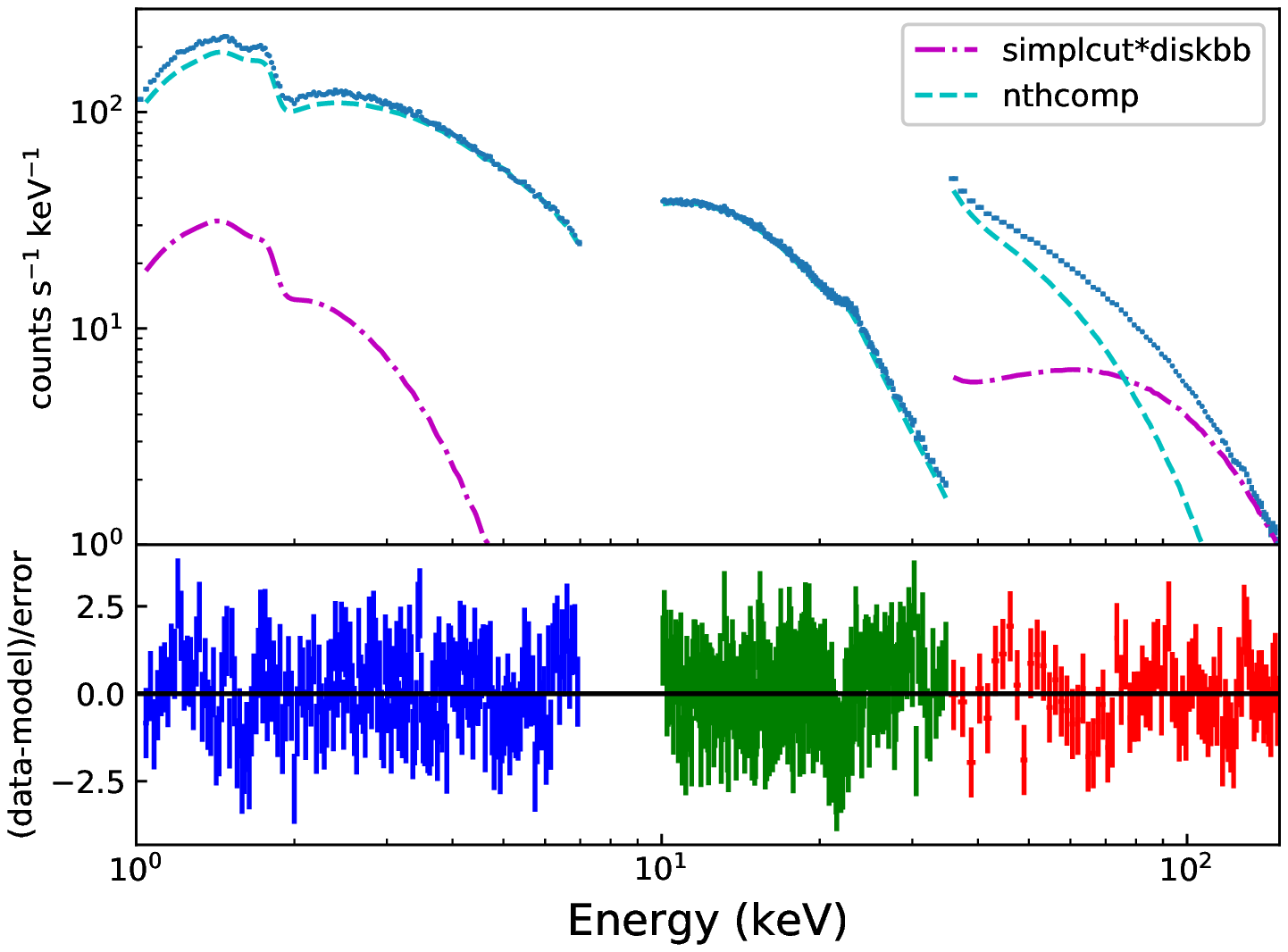}
    \caption{Top panel: spectral model and residuals for the \textit{Insight}-HXMT spectra from observation 021400200201 (first available observation in the HS), fitted with a {\tt tbabs*(diskbb+nthcomp+nthcomp)} model and free $N_{\rm H}$ (which in this case takes the value of $\approx$7.3 $ \times 10^{21}$ cm$^{-2}$).
    See Table 3 for the fit parameters. The blue datapoints in the residuals are for the LE (1.0--7.0\,keV), the green datapoints for the ME (10.0--35.0\,keV), and the red datapoints for the HE (35.0--150.0\,keV). The spectral components are described in the plot legend.
    Bottom panel: same as the top panel, but this time the model is {\tt tbabs*(simplcut*diskbb+nthcomp)} (best-fitting $N_{\rm H} \approx 7.2  \times 10^{21}$ cm$^{-2}$). See Table 4 for the fit parameters.}
    \label{fig:Comptonspec}
\end{figure}

\begin{figure}
        \includegraphics[scale=0.57,trim=20 20 0 0]{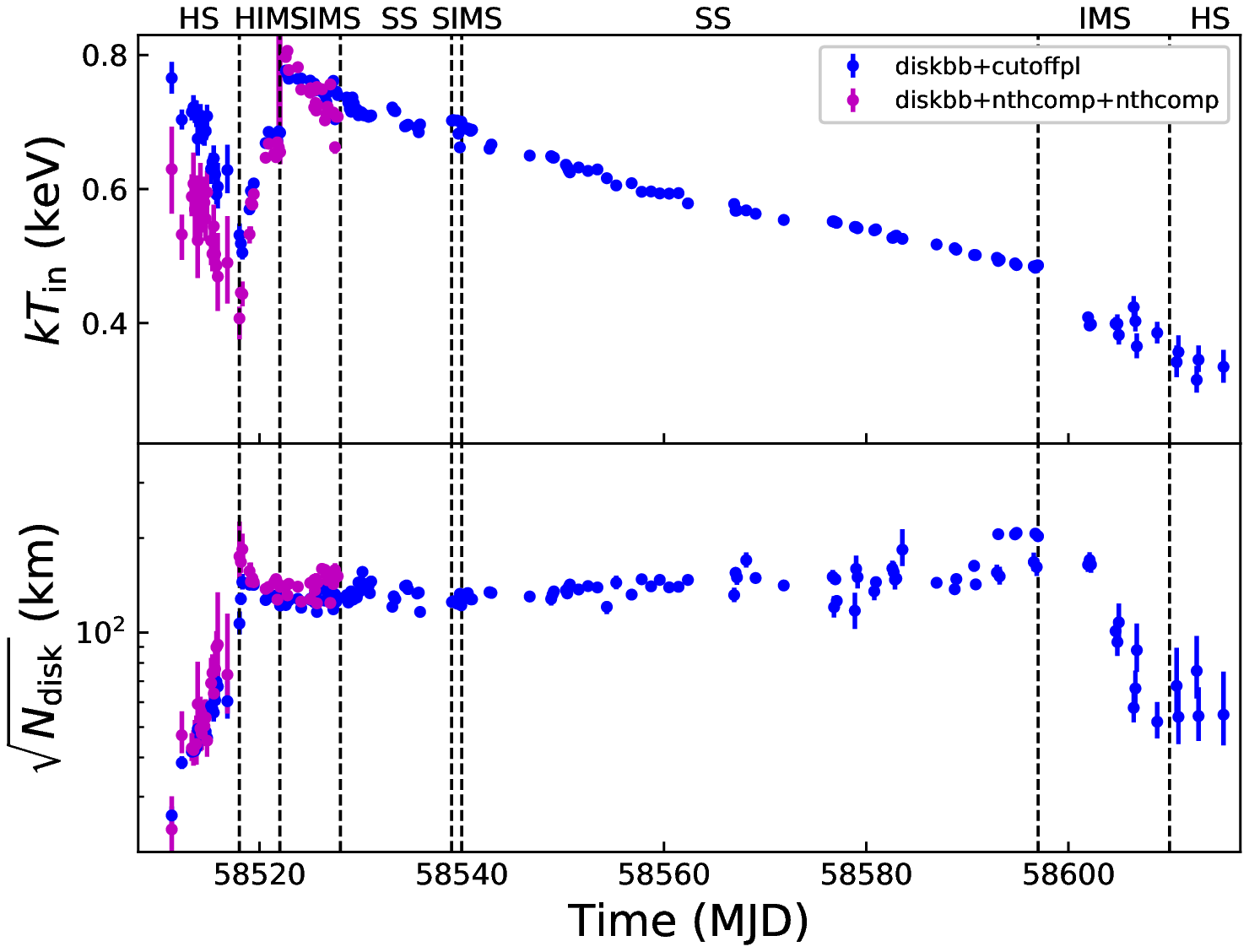}\\[-7pt]
        \includegraphics[scale=0.57,trim=20 20 0 0]{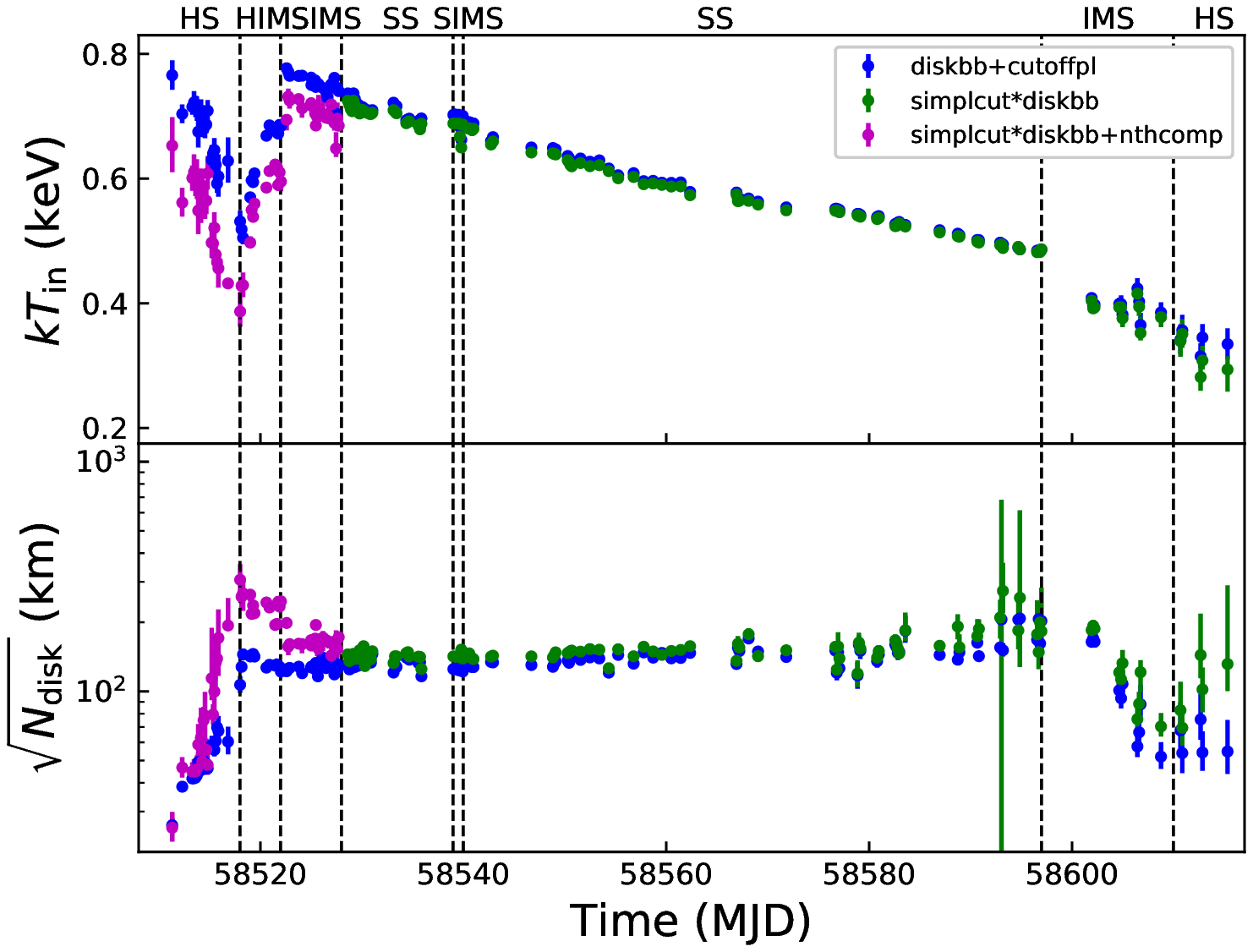}\\[-7pt]
        \includegraphics[scale=0.57,trim=20 20 0 0]{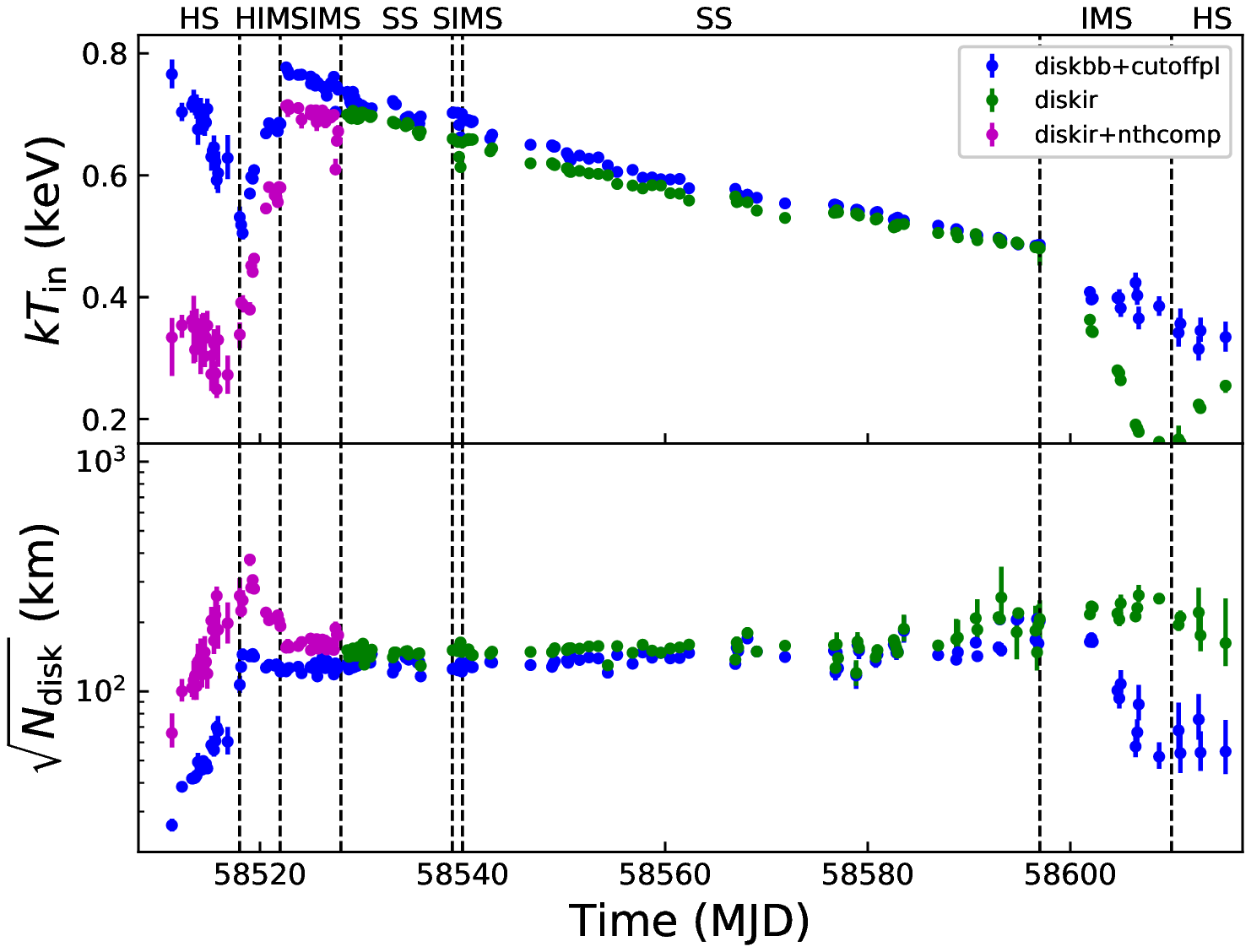}

    \caption{Disk parameter evolution in the framework of different models. In all panels, the blue datapoints are the results from {\tt diskbb+cutoffpl}. Top pair of panels: the magenta datapoints in the HS and IMS at the beginning of the outburst come from the {\tt diskbb+nthcomp+nthcomp} model. Middle pair of panels: magenta datapoints are from {\tt simplcut*diskbb+nthcomp} and green datapoints from {\tt simplcut*diskbb}. Bottom pair of panels: magenta datapoints are from {\tt diskir+nthcomp} and green datapoints from {\tt diskir}.}
    \label{fig:additonal}
\end{figure}

A simple addition of separate {\tt diskbb} and power-law or Comptonization components, as we have done so far, does not self-consistently account for the disk photons that are upscattered into the harder component(s). This loss of photons reduces the fitted normalization of the disk emission in the HS \citep{Yao2005}. As a consequence, the fitted disk radius may appear smaller than the true radius. To avoid this problem, we re-fitted the spectra with a self-consistent thermal Comptonization model, {\tt simplcut} \citep{Steiner2017}, convolved with {\tt diskbb}. We kept the scattering fraction as a free parameter in our fits, and fixed the reflection fraction to zero. We constrained the $kT_{\rm e}$ parameter to a positive value to ensure that the Compton kernel used by {\tt simplcut} is {\tt nthcomp} \citep{Steiner2017}. We also included an additional {\tt nthcomp} component in the HS and IMS. The best-fitting parameter values are summarized in Table~\ref{tab:simplcut}; peak colour temperature and normalization of the disk are shown in the middle panel of Figure~\ref{fig:additonal}; a representative spectrum is shown in Figure~\ref{fig:Comptonspec} bottom panel.  As expected, $N_{\rm disk}$ is now a factor of 2 larger in the HS and IMS (both at the beginning and at the end of the outburst) than what we had estimated previously; however, it still shows the anomalous increasing trend in the initial HS. Likewise, $T_{\rm in}$ still shows the anomalous decreasing trend in that state.

\subsubsection{{\tt diskir+nthcomp} model}
\label{sec:diskir}
As a further test, we also modelled our spectra with the Comptonization model {\tt diskir} \citep{Gierlinski2009}, with an additional {\tt nthcomp} in the HS and IMS. We fixed the values of {\it f$_{\rm in}$}, {\it r$_{\rm irr}$}, {\it f$_{\rm out}$} and $\log r_{\rm out}$ at 0.1, 1.2, 0.001 and 5, respectively; different choices of those default values do not strongly affect the other X-ray fitting parameters. 
Based on the two ranges of $kT_{\rm e}$ found in our previous fits with the double {\tt nthcomp} model (Section~\ref{sec:2nth}), we expect to find a hotter and a colder corona also with the {\tt diskir+nthcomp} model. We assumed that the hotter coronal emission is associated with {\tt diskir}, because it comes from a region closer to the inner disk, and the cooler component is modelled by {\tt nthcomp}. Accordingly, we constrained the $kT_{\rm e}$ parameter of {\tt nthcomp} to be $<$15\,keV in our fits, to avoid a fitting degeneracy with the $kT_{\rm e}$ parameter in {\tt diskir}. We find that $T_{\rm in}$ is significantly lower than estimated with the simple {\it diskbb} model (Section~\ref{sec:diskbb+po}), and $N_{\rm disk}$ is several times larger, both at the beginning and at the end of the outburst, more in agreement with what we expect from the evolution of the inner disk radius, at all stages except for the initial HS. In the initial state, we find again an anomalous increase of $N_{\rm disk}$ and a decrease of $T_{\rm in}$. However, the reduced $\chi^{2}$ values of our {\tt diskir+nthcomp} fits are much worse than those obtained from the double {\tt nthcomp} fits or from the {\tt simplcut} plus {\tt nthcomp} fits. On closer inspection, we noticed that {\tt diskir} does not reproduce well the spectral curvature above $\approx$\,50\,keV. Thus, we consider the {\tt diskir+nthcomp} parameters less reliable than those obtained with the models described in the previous sections.

\subsection{Parameter evolution assuming constant $N_{\rm H}$}
\label{sec:fixnh}
\begin{figure*}
\centering
    \includegraphics[scale=0.84,trim=25 15 20 10]{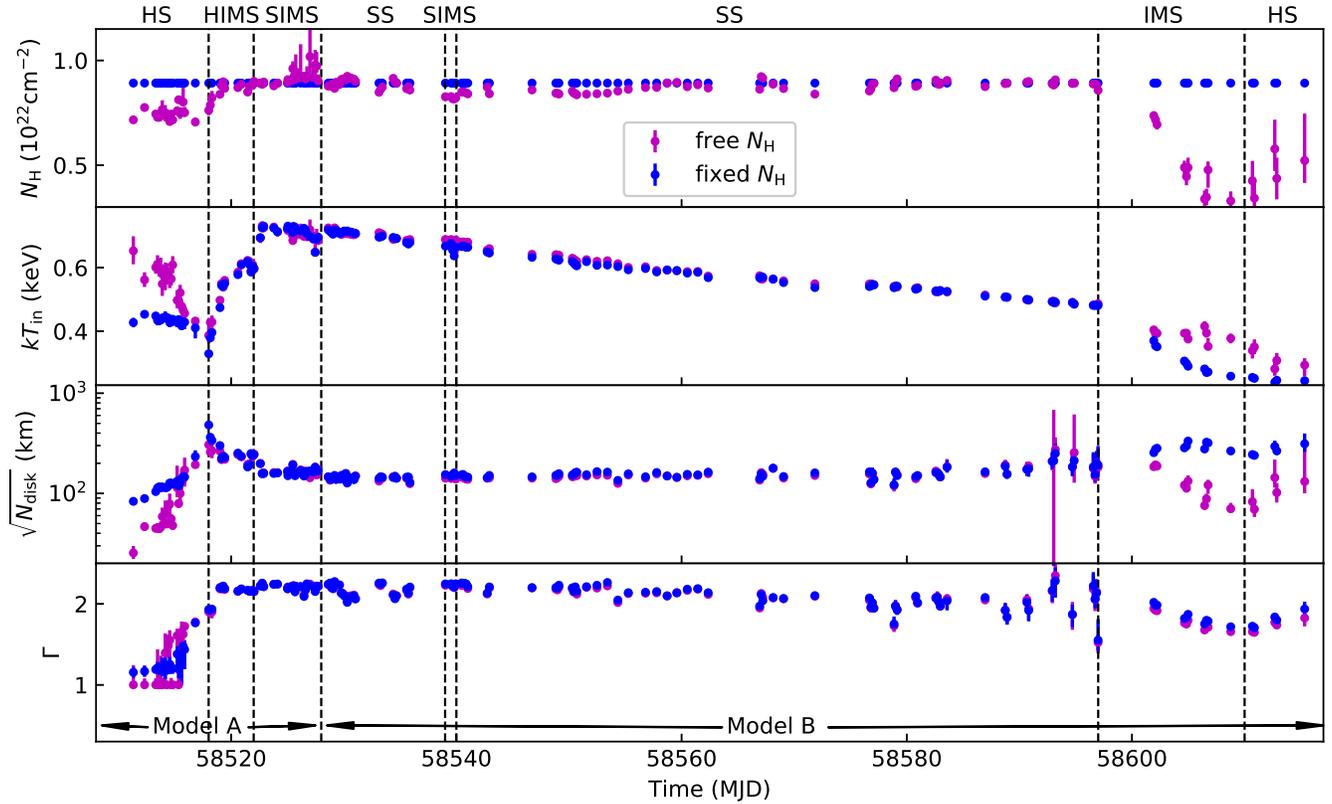}
    \caption{Evolution of $T_{\rm in}$, $\sqrt{N_{\rm disk}} (\propto r_{\rm in})$ of {\tt diskbb} and $\Gamma$ of {\tt simplcut} in the case of free $N_{\rm H}$ (magenta datapoints) and fixed $N_{\rm H}$ (blue datapoints). Model A ({\tt simplcut*diskbb+nthcomp}) is used for all spectra before the first transition to the SS; Model B ({\tt simplcut*diskbb}) is used for all spectra subsequent to that transition. $N_{\rm H}$ is fixed at $8.9 \times \rm 10^{21}\ cm^{-2}$.}
    \label{fig:fixednh}
\end{figure*}

\begin{figure*}
\centering
        \includegraphics[scale=0.58,trim=20 10 10 20]{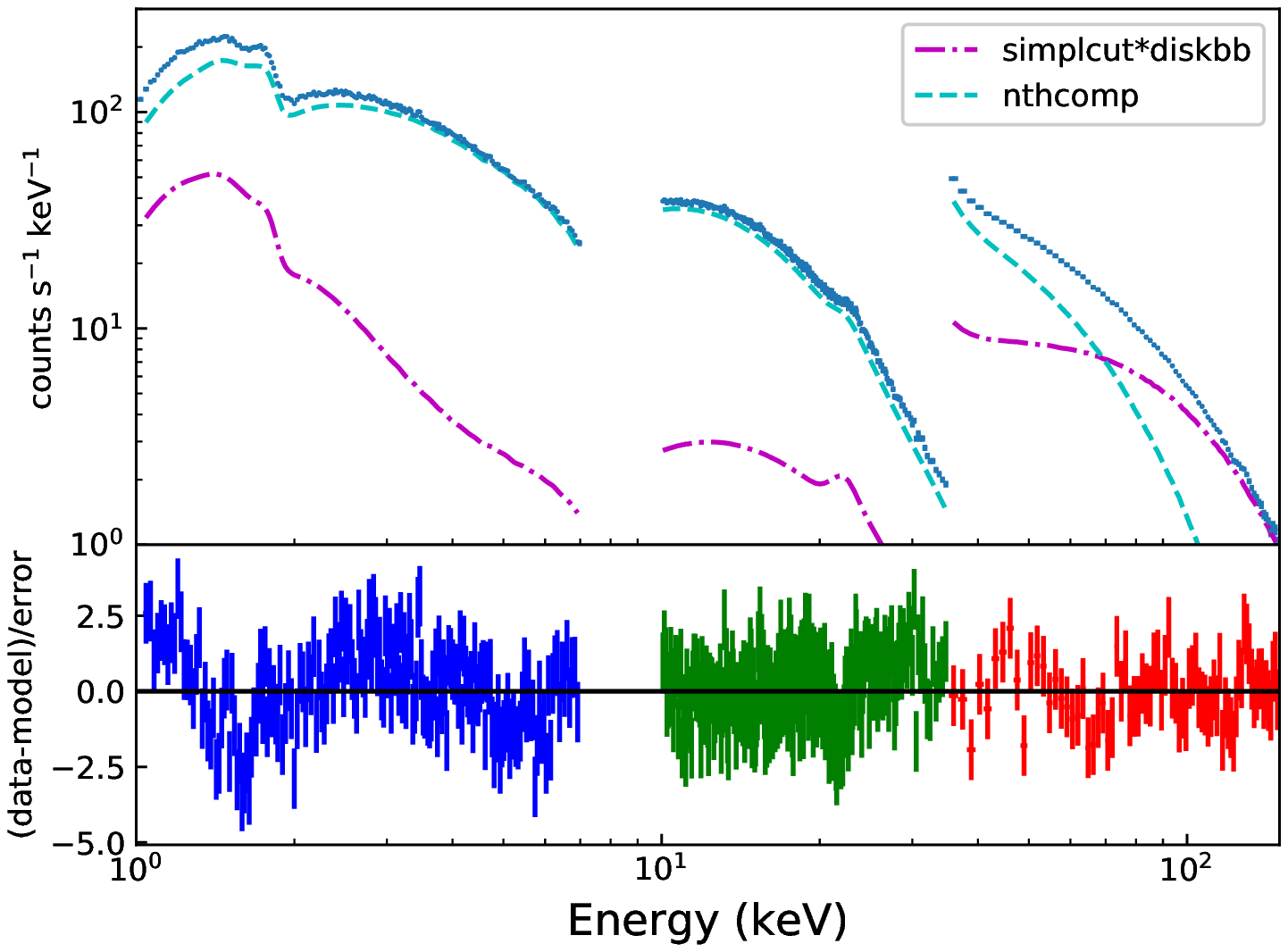}
        \includegraphics[scale=0.58,trim=10 10 20 20]{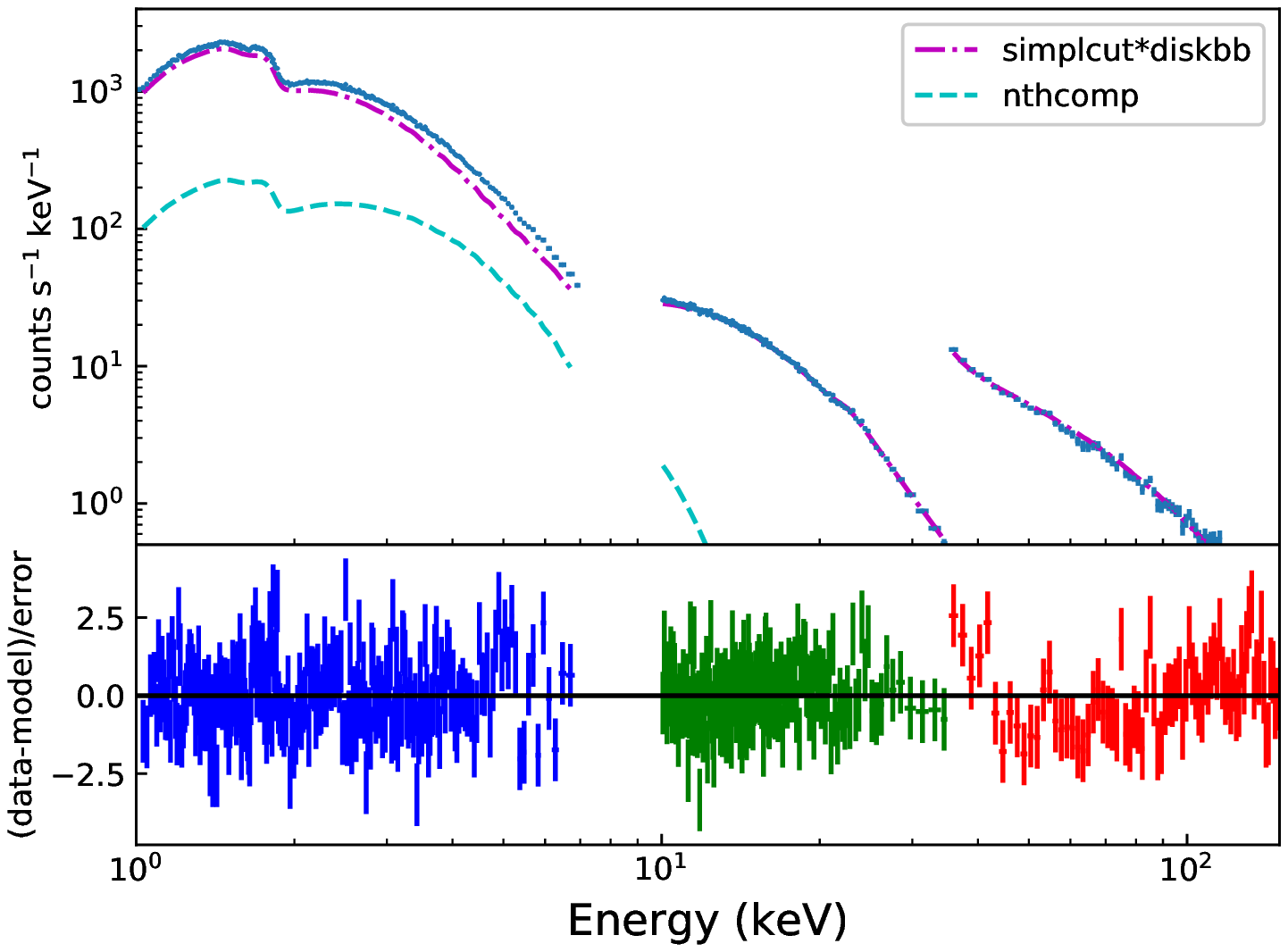}

    \caption{Left panel: spectral model and residuals for the \textit{Insight}-HXMT spectra from observation 021400200201 (first available observation in the HS), fitted with a {\tt tbabs*(simplcut*diskbb+nthcomp)} model and $N_{\rm H}$ fixed to the median value in the SIMS ($N_{\rm H} = 8.9 \times \rm 10^{21}\ cm ^{-2}$). The fit in the soft X-ray band is not as good as the one obtained with free (lower) $N_{\rm H}$, shown in the bottom panel of Figure~\ref{fig:Comptonspec}. Colors and symbols are the same as in Figure~\ref{fig:Comptonspec}. 
    Right panel: same as in the left panel, for the spectrum from observation 021400201701 (SIMS). Here the fixed value of $N_{\rm H} = 8.9 \times \rm 10^{21}\ cm ^{-2}$ provides an excellent fit to the soft band.}
    \label{fig:simplspec}
\end{figure*}

In all the modelling described so far, we have always kept $N_{\rm H}$ as a free parameter. Its best-fitting value remains approximately constant throughout the SIMS and SS, but is lower in the initial and final HSs. We cannot rule out that this is a physical effect: for example, accretion disk winds may contribute to the absorbing column in the SS. However, more plausibly, the apparent change in $N_{\rm H}$ is only a fitting artifact, related to a degeneracy in the fit parameters. Thus, we need to study how the other parameters vary when we assume a constant $N_{\rm H}$. For this analysis, we selected the {\tt {simplcut*diskbb+nthcomp}} model (Section~\ref{sec:simplcut}) as the most physical choice, based on our detailed comparison of alternative models (Sections ~\ref{sec:diskbb+po}--~\ref{sec:diskir}). We determined the median value of $N_{\rm H}$ during the SIMS, $N_{\rm H} \approx\,8.9 \times 10^{21}$ cm$^{-2}$. We froze $N_{\rm H}$ at that value for every epoch. We also fixed the seed photon temperature of {\tt nthcomp} at 0.1 keV; we verified that the parameter evolution trends do not change if we link the seed photon temperature instead to $T_{\rm in}$ of {\tt diskbb}. The best fitting parameter values are shown in Figure~\ref{fig:fixednh}. Two representative spectra, one from the HS and one from the SIMS, are shown in Figure~\ref{fig:simplspec}.

The only significant differences between the free-$N_{\rm H}$ and fixed-$N_{\rm H}$ are in the initial and final stages of the outburst. In the initial HS, the peak colour temperature decreases from $\approx$\,0.45 keV to $\approx$\,0.35 keV. This decline is less dramatic than we found from the free-$N_{\rm H}$ fit, but it is still significant, at a time when the canonical outburst evolution predicts a temperature increase. The temperature trend reverses and becomes consistent with a standard disk evolution from the beginning of the HIMS. At the same time, the apparent inner radius $r_{\rm in}$ increases by a factor of 2 in the initial HS---a more moderate increase than what we found with a free $N_{\rm H}$, but still significant, in a stage where $r_{\rm in}$ is supposed to decrease. Again, the trend reverses from the beginning of the HIMS. The photon index of the comptonized component increases during the initial HS but remains $\lesssim$\,2 until the beginning of the HIMS; its behaviour is not substantially affected by the choice of $N_{\rm H}$.

In the IMS and HS at the end of the outburst, the choice of a fixed $N_{\rm H}$ leads to a moderate increase in the apparent radius by a factor of $\approx$\,1.5 (Figure~\ref{fig:fixednh}), associated with a decrease in colour temperature. This trend is consistent with the canonical evolution, and removes the contradictory behaviour (radius and temperature simultaneously decreasing) previously inferred in our free-$N_{\rm H}$ modelling.

In summary, the results of our modelling with fixed $N_{\rm H}$ showed a canonical end of the outburst but an anomalous evolution of the disk component in the hard rising phase. We verified that the same trend is found with the other models previously considered (Sections~\ref{sec:diskbb+po},~\ref{sec:2nth},~\ref{sec:diskir}), when we refit them at fixed $N_{\rm H}$. We will propose a physical interpretation for the anomalous initial phase in Section~\ref{sec:dis}.

\section{Discussion}
\label{sec:dis}

Based on the hardness ratios (Figure~\ref{fig:lc}), the HIDs (Figure~\ref{fig:hids}) and the spectral parameters (Figure~\ref{fig:spec}), we have shown that MAXI J1348$-$630 completed a typical BHB outburst evolution cycle from MJD 58510 to MJD 58617. It was detected first in the HS, went through a hard-to-soft transition to reach the SS, before returning to the HS.

\begin{figure*}
    \centering
    \subfigure{
        \begin{minipage}{8.8cm}
        \centering
        \includegraphics[scale=0.56,trim=20 10 0 0]{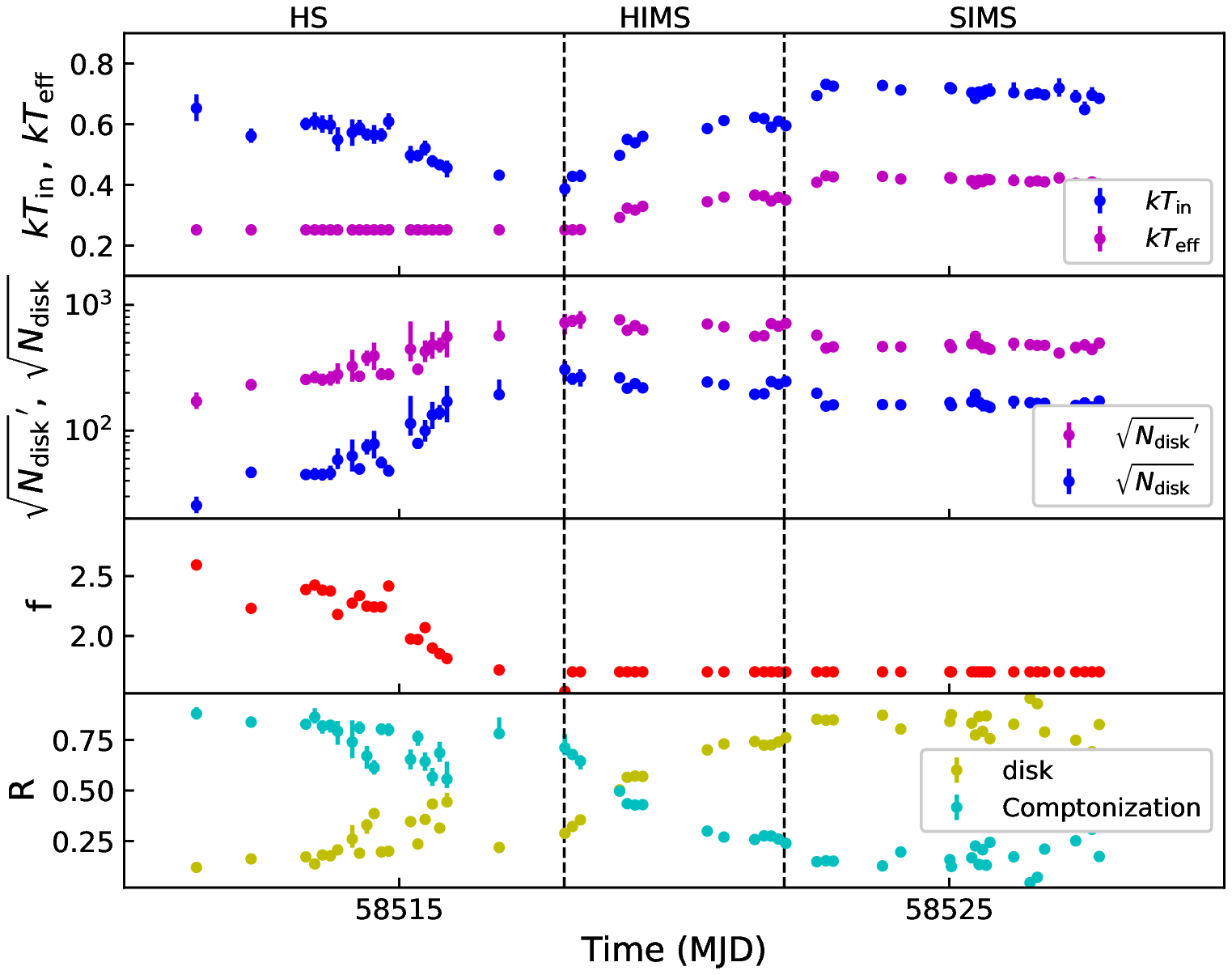}
        \end{minipage}}
    \subfigure{
        \begin{minipage}{8.8cm}
        \centering
        \includegraphics[scale=0.56,trim=10 10 0 0]{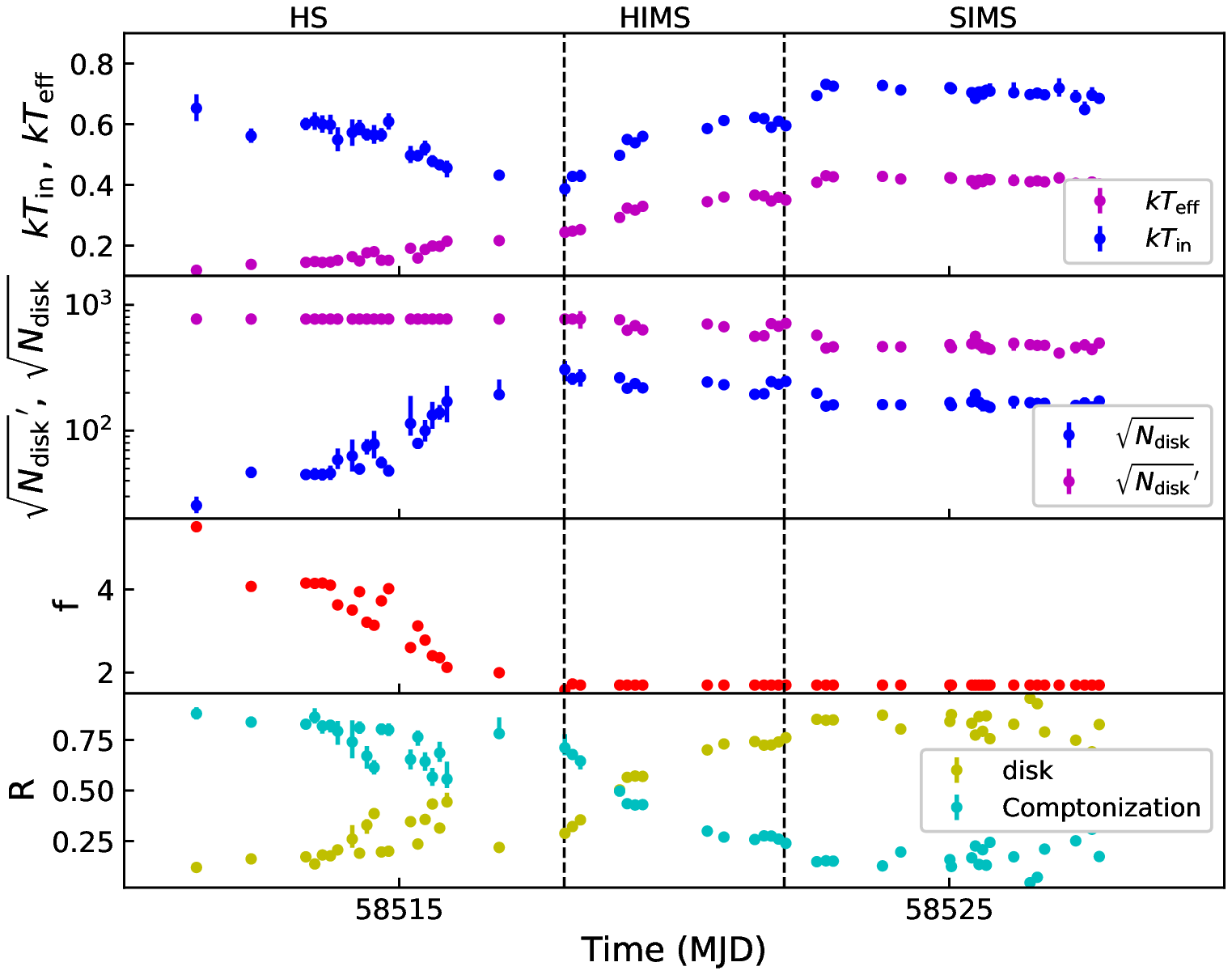}
        \end{minipage}}
    \\
    \subfigure{
        \begin{minipage}{8.8cm}
        \centering
        \includegraphics[scale=0.56,trim=20 10 0 0]{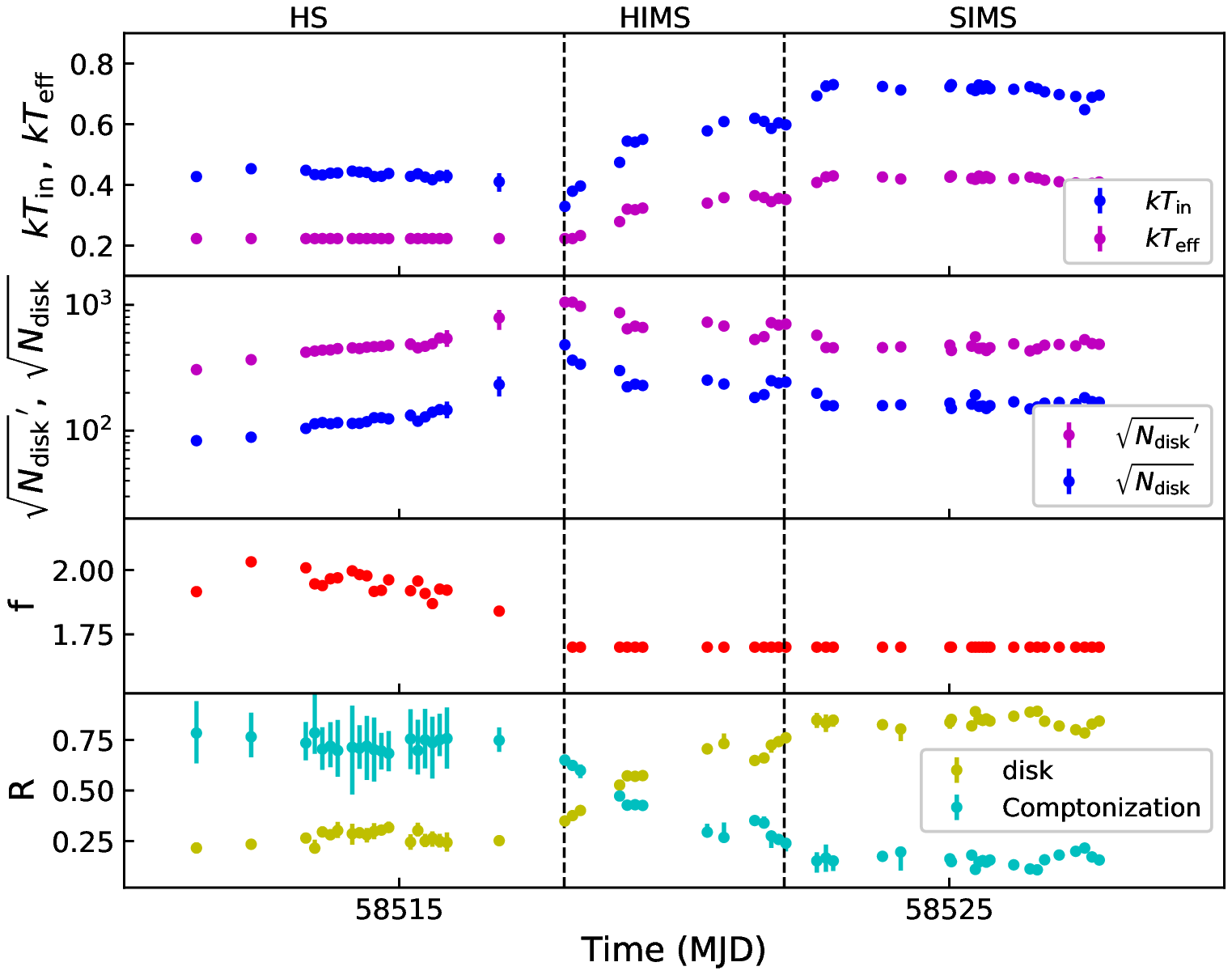}
        \end{minipage}}
    \subfigure{
        \begin{minipage}{8.8cm}
        \centering
        \includegraphics[scale=0.56,trim=10 10 0 0]{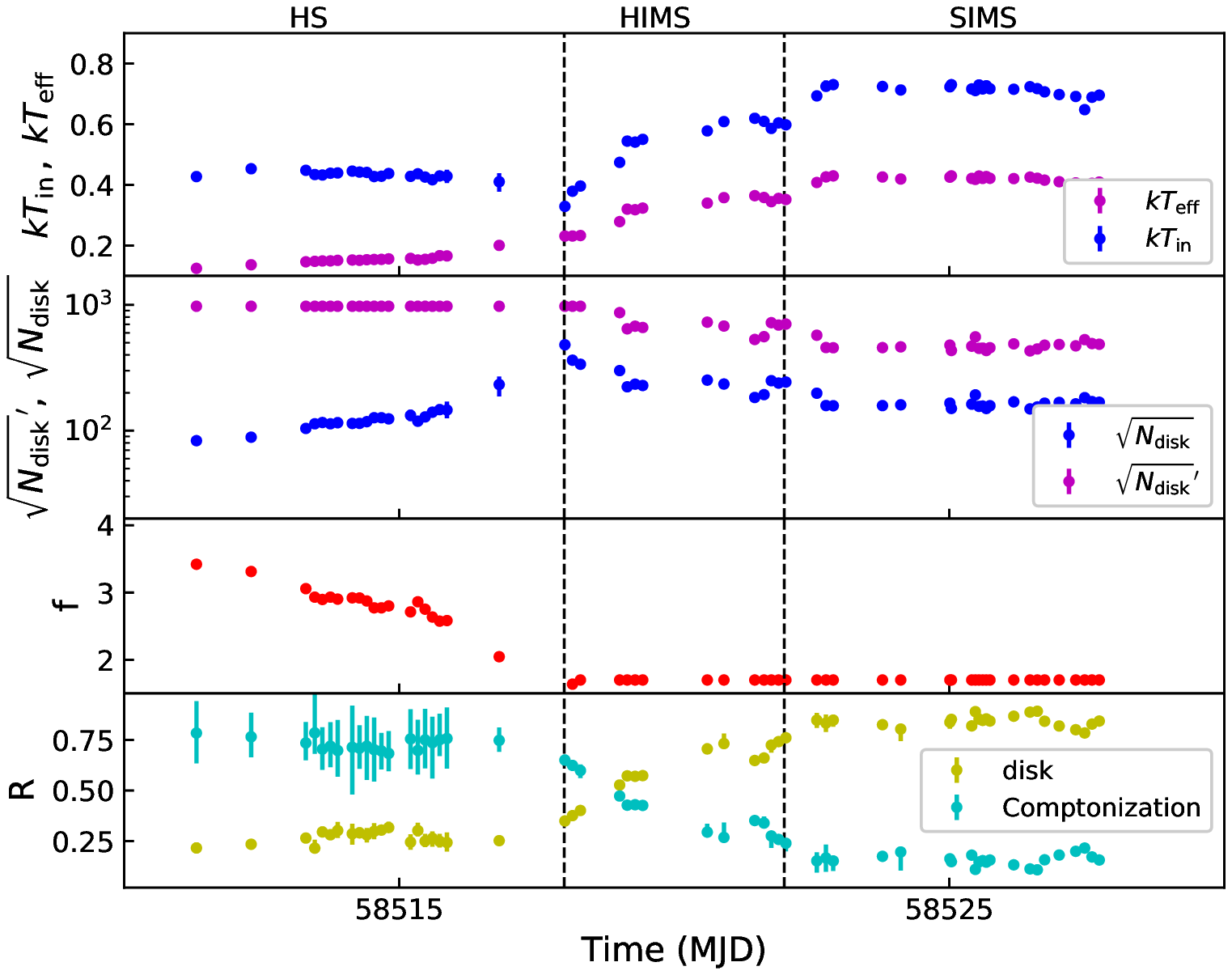}
        \end{minipage}}    
    \caption{
    Comparison between apparent and physical evolution of disk temperature and radii, as a function of hardening factor $f$, as well as the evolution of the bolometric flux percentage (R) for the {\tt diskbb} and {\tt simplcut} components. Top left panel: blue datapoints represent the apparent evolution ($kT_{\rm in}$ (keV), $\sqrt{N_{\rm disk}}$ (km)) in the model described in Section~\ref{sec:simplcut}, for free $N_{\rm H}$. Magenta datapoints represent the physical evolution ($kT_{\rm eff}$ (keV), $\sqrt{N_{\rm disk}}'$ (km)), if $f$ is chosen in such a way to keep $kT_{\rm eff}$ constant in the initial HS. The evolution of $f$ is also plotted (red datapoints). Top right panel: as in the top left panel, but this time $f$ is defined in a way to keep $\sqrt{N_{\rm disk}}'$ constant in the HS. Bottom left panel: as in the top left panel, but for fixed-$N_{\rm H}$ fitting (Section~\ref{sec:fixnh}). Bottom right panel: as in the top right panel, but for fixed-$N_{\rm H}$ fitting. We suggest (Section~\ref{sec:con}) that the bottom right panel represents the most physical conditions, as $kT_{\rm eff}$ is now increasing during outburst rise, and $f$ is decreasing smoothly from $\approx$\,3.5 to the canonical value of $\approx$\,1.7.}
    \label{fig:harden}
\end{figure*}

A soft component consistent with a thermal model is significantly detected in our spectral fitting at all epochs, including in the initial HS, although it is only contributing a relatively small fraction of flux in that state, compared with the power-law component (Figure~\ref{fig:harden}).
What is the origin of this component, especially in the HS? The question is still open. \citet{2001ApJ...554..528N} argued that the soft variable component of Cyg X-1 in the HS can be explained by highly variable X-ray shots. Alternatively, from wide-band \textit{Suzaku} observations of Cyg X-1, \citet{2013PASJ...65...80Y} showed that in addition to two Comptonized components there are two types of low energy components seen around 1 keV: a variable soft component and the Wien tail of a static disk. For MAXI J1348$-$630, we argue that this component always comes from the disk, throughout the outburst. The reason we suggest this interpretation is that its model parameters evolve smoothly and continuously from the initial HS to the SS, when it becomes the dominant component. There is little doubt (also by analogy with other BHBs) that the thermal component in the SS is associated with an optically thick disk, extending down to ISCO. Before then, during the outburst evolution from HS to SS, we do not see any transition or discontinuity that could suggest the appearance of the disk component and the disappearance of a different source of soft emission. Thus, the disk scenario is the simplest explanation consistent with the data, although we cannot rule out alternative ones.

As we pointed out earlier, MAXI J1348$-$630 shows two physically distinct Comptonization components. The {\tt nthcomp} component (which we may interpret as the classical corona) dominates the broadband spectrum at the beginning of the outburst; the {\tt simplcut*diskbb} component dominates the SIMS and SS. Physically, this means that the disk does not reach ISCO in the initial HS, and the comptonized disk emission cannot be the dominant component in that state. Later, during the SIMS and SS, the disk radius settles at a constant value consistent with ISCO, and dominates the emission.

As outlined so far, this outburst scenario may appear similar to any other canonical outburst. However, this is not the case. The best-fitting disk parameters show a peculiar behavior in the initial HS. $N_{\rm disk}$ starts out even smaller than in the SS, then increases until it reaches a peak (whose value is model dependent), and finally decreases (as expected) until the disk reaches ISCO. The reason this is apparently unphysical is that if the inner disk radius was really as small or smaller than ISCO in the HS, the thermal component would already dominate the X-ray emission at that stage (which is not the case). Similarly, $T_{\rm in}$ starts out much hotter than expected for a truncated hard-state disk \citep[e.g.,][]{2004MNRAS.355.1105F, 2006ARA&A..44...49R, Done2007}, then decreases towards more plausible values (also model-dependent and very much coupled with the $N_{\rm disk}$ behaviour), and finally increases until it reaches the expected $L_{\rm X}$ vs $T_{\rm in}^4$ relation. A (model-dependent) irregular behaviour is also observed in the HS at the end of the outburst, with $N_{\rm disk}$ either constant or decreasing again, and $T_{\rm in}$ either constant or slightly increasing, at odds with the expectations of a retreating inner disk.

Most BHB outbursts follow the canonical evolution: large and decreasing $N_{\rm disk}$, low and increasing $T_{\rm in}$ in the initial HS, then a constant $N_{\rm disk}$. For example, MAXI J1659$-$152 \citep{Munoz-Darias2011, Yamaoka2012},  XTE J1550$-$564 \citep{Rodriguez2003}, XTE J1908$+$094 \citep{Tao2015,Zhang2015} and MAXI J1535$-$571 \citep{Tao2018}. However, the anomalous behaviour seen in the initial HS of MAXI J1348$-$630 is far from rare (even though it is rarely discussed). Previous examples include several outbursts of GX 339-4 \citep{Dunn2008,Nandi2012}, of GRO J1655$-$40 \citep{Done2007,Debnath2008}, of XTE J1650$-$500 \citep{Gierlinski2004} and a few other cases illustrated by \cite{2011MNRAS.411..337D}.

A partial explanation for the initial low value of $N_{\rm disk}$ is that a significant fraction of the seed disk photons are upscattered into the power-law component \citep{Yao2005}. This effect is accounted for when we replaced the independent disk-blackbody plus power-law components with the self-consistent Comptonization models {\tt diskir} and {\tt simplcut}*{\tt diskbb}. Those models provide the intrinsic disk normalization (and, hence, inner radius) before Compton scattering. However, even with those two models, $N_{\rm disk}$ is smaller than expected in the initial HS (in particular, smaller than in the SS). The difference between the best-fitting $N_{\rm disk}$ from {\tt diskir} and {\tt simplcut}*{\tt diskbb} comes from the different Compton kernel in the two models, and from the effect of coronal irradiation of the disk (included in {\tt diskir} but not in {\tt simplcut}*{\tt diskbb}). Simulations suggest \citep{Yao2005} that the detailed geometry of the corona affects the apparent disk flux: an incorrect approximation for the scattering region may be the reason why $N_{\rm disk}$ appears too small. Quantitative discussions on how $N_{\rm disk}$ depends on the geometry of the corona is beyond the scope of this work.

However, the fact that the anomalous behavior appears simultaneously in $N_{\rm disk}$ and $T_{\rm in}$, and that the two parameters are strongly coupled, suggests that additional processes are in play, such as changes in color correction factor $f$ in different states (originally proposed by \citealt{Soria2008} and \citealt{2011MNRAS.411..337D}). The fitted peak color temperature $T_{\rm in}$ is equal to $f T_{\rm eff}$, where $T_{\rm eff}$ is the peak effective temperature ({\it i.e.}, a more physical measure of the true disk emission). Conversely, the fitted normalization $N_{\rm disk} \propto r_{\rm in}^2 \propto R_{\rm in}^2/f^4$, where $r_{\rm in}$ is the apparent radius, and $R_{\rm in}$ is the physical radius. The canonical value of the hardening factor is $f \approx\,1.7$ \citep{Shimura1995, Shafee2006}. The change of $f$ (for any reason) would lead to simultaneous variations of both $T_{\rm in}$ and $N_{\rm disk}$ (even after the correction of the lost soft photons in inverse Compton scattering), and may drive the observed anomalous behaviors. Hardening factors as high as $\approx$3 in the initial phase of an outburst were suggested by \cite{2011MNRAS.411..337D} to explain the anomalously low values of $R_{\rm in}$ in some of the BH transients in their sample.

We tested this hypothesis for MAXI J1348$-$630 by estimating what values of $f$ are needed to make the apparent temperature and radius evolution in the initial HS consistent with their expected physical evolution (Figure~\ref{fig:harden}). First, we assumed that $T_{\rm eff}$ is constant in that state, then we assumed that $R_{\rm in}$ is constant. In both cases, the inferred values of $f$ are to be taken as lower limits, as even higher values would be needed to produce an increasing $T_{\rm eff}$ or a decreasing $R_{\rm in}$. We repeated this exercise first with the set of model parameters derived with free $N_{\rm H}$ (Section~\ref{sec:2nth}), and then with the parameters derived with fixed $N_{\rm H}$ (Section~\ref{sec:fixnh}). The more regular evolution of the $f$ parameter in the fixed-$N_{\rm H}$ scenario (Figure~\ref{fig:harden}, bottom panels) is an indication that this constraint is preferable to the free-$N_{\rm H}$ scenario (Figure~\ref{fig:harden}, upper panels); however, the qualitative trend is the same in the two cases. The values of $f$ quoted below refer to the fixed-$N_{\rm H}$ case. Imposing a constant $T_{\rm eff}$ in the initial HS requires $f \approx\,2$; however, this is not high enough to offset the anomalous behaviour of the inner radius, as $R_{\rm in}$ is still unphysically increasing (Figure~\ref{fig:harden}, bottom left panel). By contract, imposing a constant $R_{\rm in}$ in the HS requires an initial $f \approx\,3.5$, declining smoothly to $f \approx\,1.7$ in the HIMS (Figure~\ref{fig:harden}, bottom right panel). With this choice of $f$, the anomalous temperature evolution is also normalized, as $T_{\rm eff}$ now increases with time during outburst rise. The subsequent evolution in the HIMS can be described as the continuing increase of $T_{\rm eff}$ and decrease of $R_{\rm in}$ down to ISCO at constant $f$, in agreement with the standard outburst model. In this scenario, the transition between HS and HIMS can be defined as the time when the incipient inner disk (condensing from the cooling corona) has reached a sufficiently high effective optical thickness to be described by the standard Shakura-Sunyaev solution. The transition between HIMS and SIMS is when the optically thick inner disk has reached ISCO.

What could produce the initial high value of $f$, later declining to the standard value? One of the physical parameters that determine the value of $f$ are the relative values of free-free absorption and electron scattering coefficients. From textbook radiation theory \citep{1979rpa..book.....R}, we note (see also \citealt{Soria2008}) that in the region of the disk where the opacity is dominated by electron scattering rather than free-free absorption, the emerging flux at frequency $\nu$ is reduced by a factor $\left(\alpha^{\rm ff}_{\nu}/\alpha^{\rm es}   \right)^{1/2} < 1$, where $\alpha^{\rm ff}_{\nu}$ is the free-free absorption coefficient and $\alpha^{\rm es}$ is the electron scattering coefficient. If we impose that the disk is radiatively efficient, {\it i.e.} that it dissipates all the gravitational energy without advection, the color temperature of the emerging radiation must increase by a factor $\left(\alpha^{\rm es}/\alpha^{\rm ff}_{\nu}   \right)^{1/8} > 1$ compared with the effective temperature of blackbody emission with the same flux. From the standard disk solution \citep{1973A&A....24..337S, 2002apa..book.....F} and the expression for the scattering and free-free absorption coefficients, it follows that the hardening factor $f \sim \left(\alpha^{\rm es}/\alpha^{\rm ff}_{\nu} \right)^{1/8} \propto n_e^{-1/8}$, where $n_e$ is the electron density. For a disk around a 10-$M_{\odot}$ BH, with an accretion rate $\approx\,0.1$ Eddington, at a radius $\approx$\,10 times $R_{\rm ISCO}$, the characteristic electron density in equilibrium is $n_e \sim 10^{22}$ cm$^{-3}$. But the inner disk around  MAXI J1348$-$630 was far from equilibrium (for a given accretion rate) at the start of the outburst. We speculate that it was still in the process of condensing from the hot corona \citep{Liu2007,Taam2008,Meyer-Hofmeister2009,Liu2011,Qiao2017}. Although already optically thick and radiatively efficient, its density may have been a few orders of magnitude below the equilibrium value. An increase in $n_e$ by three orders of magnitudes corresponds to a decrease of $f$ by a factor of 2, which would explain the initial odd behaviour of disk normalization and color temperature. We further speculate that the transition to the HIMS corresponds to the moment when the inner disk structure reaches the equilibrium configuration. Other effects that determine the value of $f$ include Comptonization near the disk surface \citep{Shimura1993}: we expect that a cooling/collapse of the hot corona during the evolution from harder to softer states will reduce the amount of up-scattering suffered by emerging disk photons and therefore also reduce the value of $f$. More detailed modelling of such evolution is beyond the scope of this work.

\section{Conclusions}
\label{sec:con}
We presented the X-ray spectral behaviour of the BH candidate MAXI J1348$-$630 during its main outburst in 2019, based on \textit{Insight}-HXMT and \textit{Swift} observations. The source was discovered in the HS, then transited to the SS, and finally returned to the HS, consistent with the typical outburst of BHBs. Throughout the outburst, the spectra are well described by a model consisting of a disk component and either one or two power-law-like components. In the SIMS and SS, the disk properties are consistent with the standard model ($R_{\rm in} \approx\,R_{\rm isco}$, $L_{\rm disk} \propto T_{\rm in}^4$). Instead, in the initial HS (first week of \textit{Insight}-HXMT observations), the disk behaviour appears odd, with a increasing disk normalization and a decreasing color temperature. The canonical model of outburst evolution predicts an initial decrease of the disk normalization, as the inner disk radius moves closer to ISCO, and a corresponding temperature increase.

We showed that part of the apparent initial deficit of photons in the disk normalization is caused by the upscattering of thermal photons into the Comptonized component. This is why self-consistent Comptonization models such as {\tt diskir} and {\tt simplcut}*{\tt diskbb} are preferable to the simpler disk-blackbody plus power-law model, when a corona is still present. However, even after accounting for such photon transfer, the initial odd trend of $N_{\rm disk}$ and $T_{\rm in}$ persists. We showed that the coupled evolution of those two quantities can be phenomenologically described with an effective hardening factor about 2 times higher at the start of the observations than in the rest of the outburst (where we assume a canonical value of 1.7). We argued that this change in hardening factor reveals a real and perhaps rarely observed phase of disk evolution. A higher-than-usual color correction is theoretically expected from an optically thick, radiatively efficient disk annulus (in the region where scattering opacity dominates) if the electron density is much lower than the equilibrium value (given by the solutions of the standard disk equation). We speculated that at the start of our observations, the inner part of the standard thin disk was still in the process of condensing from the previous geometrically thick hot corona. Its density was still increasing (implying a decreasing hardening factor) until it reached the equilibrium value; this happened as the system switched from the HS to the HIMS.

In conclusion, the outburst rise in BHBs is often empirically simplified as the inward motion of the inner truncation radius of a standard disk, replacing the geometrically thick, hot scattering region (with a well defined, sharp boundary between the two regions). In this system, and perhaps in others, we may have seen instead the gradual condensation of the hot scattering region into a thin standard disk, during the initial HS. Once this part of the disk reached its equilibrium, we also saw a canonical evolution for a few days (the short transition phase identified as HIMS), with the inner radius moving further inwards until it reached ISCO, where it remained throughout the SS.

\acknowledgments
This work made use of data from the Insight-HXMT mission, a project funded by China National Space Administration (CNSA) and the Chinese Academy of Sciences (CAS). This work is supported by the National Key R\&D Program of China (2021YFA0718500). We are grateful to the anonymous referee for helpful comments and suggestions. We acknowledges funding support from the National Natural Science Foundation of China (NSFC) under grant Nos. 12073029, 12122306, U1838115, U1838108, U1838201, U1838202, U1938104, U2031205, U2038103 and 11733009, the CAS Pioneer Hundred Talent Program Y8291130K2 and the Scientific and technological innovation project of IHEP Y7515570U1.
{\it Facilities:} \textit{Insight}-HXMT, \textit{Swift}

\clearpage

\LongTables 
\begin{deluxetable}{lcccccccccc}
\tablecolumns{10}
\tablewidth{0pc}
\tablewidth{\textwidth}
\tabletypesize{\scriptsize}
\tablecaption{Log of the \textit{Insight}-HXMT observations, and best-fitting parameters when the spectra are modelled with [{\tt diskbb+cutoffpl}].\\  Error ranges represent the 90\% confidence limit for one interesting parameter, with the $\chi^2$ fitting statistics.\\  HS – hard state; HIMS – hard intermediate state; SIMS – soft intermediate state; SS – soft state.  \label{tab:HXMTobs}}
\tablehead{
\colhead{ObsID}  & \colhead{Observed date} & \colhead{Exposure time} & \colhead{ExposureID}  & \colhead{$N_{\rm H}$} & \colhead{$kT_{\rm in}$} & \colhead{$N_{\rm disk}$} & \colhead{$\Gamma$} & \colhead{$E_{\rm cut}$} & \colhead{State} \\
 \colhead{}  & \colhead{(day)} & \colhead{(ks)} & \colhead{} & \colhead{($\rm 10^{22}\ cm^{-2}$)} & \colhead{(keV)} & \colhead{} & \colhead{} & \colhead{(keV)}
}
\startdata
\multirow{1}{*}{0214002002} & \multirow{1}{*}{2019-01-28 07:31:08} & \multirow{1}{*}{10}  & 021400200201 & $0.607^{+0.017}_{-0.017}$   & $0.77^{+0.02}_{-0.02}$       & $680^{+90}_{-80}$       & $1.302^{+0.010}_{-0.010}$ & $86.6^{+1.3}_{-1.3}$ & \multirow{1}{*}{HS} \\
\noalign{\smallskip}\hline\noalign{\smallskip}
\multirow{1}{*}{0214002003} & \multirow{1}{*}{2019-01-29 07:22:59} & \multirow{1}{*}{10}  & 021400200301 & $0.653^{+0.014}_{-0.014}$   & $0.704^{+0.015}_{-0.015}$    & $1480^{+150}_{-130}$    & $1.296^{+0.007}_{-0.007}$ & $72.9^{+0.7}_{-0.7}$ &\multirow{1}{*}{HS} \\
\noalign{\smallskip}\hline\noalign{\smallskip}
\multirow{17}{*}{0214002004} & \multirow{17}{*}{2019-01-30 07:15:00} & \multirow{17}{*}{130} & 021400200401 & $0.632^{+0.011}_{-0.011}$   & $0.715^{+0.013}_{-0.012}$    & $1740^{+140}_{-130}$    & $1.333^{+0.006}_{-0.006}$ & $68.4^{+0.6}_{-0.6}$ &\multirow{17}{*}{HS} \\
& & & 021400200402 & $0.614^{+0.016}_{-0.016}$   & $0.722^{+0.018}_{-0.018}$    & $1750^{+190}_{-170}$    & $1.313^{+0.013}_{-0.013}$ & $67^{+2}_{-2}$       & & \\
& & & 021400200403 & $0.617^{+0.015}_{-0.014}$   & $0.715^{+0.017}_{-0.017}$    & $1770^{+190}_{-170}$    & $1.347^{+0.006}_{-0.006}$ & $68.3^{+0.6}_{-0.6}$ & & \\
& & & 021400200404 & $0.624^{+0.017}_{-0.017}$   & $0.712^{+0.020}_{-0.019}$    & $1900^{+200}_{-200}$    & $1.356^{+0.007}_{-0.007}$ & $68.4^{+0.6}_{-0.6}$ & & \\
& & & 021400200405 & $0.65^{+0.03}_{-0.03}$      & $0.68^{+0.03}_{-0.03}$       & $2400^{+500}_{-400}$    & $1.358^{+0.007}_{-0.007}$ & $67.9^{+0.6}_{-0.6}$ & & \\
& & & 021400200407 & $0.64^{+0.03}_{-0.02}$      & $0.70^{+0.03}_{-0.03}$       & $2200^{+400}_{-300}$    & $1.343^{+0.009}_{-0.009}$ & $65.1^{+0.9}_{-0.9}$ & & \\
\enddata
\end{deluxetable}
\footnotesize{\textbf{Note.--}
$N_{\rm H}$ is the absorption column density in units of $10^{22}~\rm{atoms~cm^{-2}}$; $kT_{\rm in}$ is the inner disk color temperature in units of keV; $N_{\rm disk}$ is the normalization of the {\tt diskbb} model; $\Gamma$ is the photon index of the {\tt cutoffpl} model; $E_{\rm cut}$ is the e-folding energy of exponential rolloff in units of keV.} \\
\footnotesize{$^a$ The error reaches the limit of parameter.}

\begin{deluxetable}{lcccccccc}
\tablecolumns{8}
\tablewidth{0pc}
\linespread{1.6}
\tablecaption{Log of the \textit{Swift}/XRT observations, and best-fitting parameters when the spectra are modelled with [{\tt diskbb+powerlaw}].\\ Error ranges represent the 90\% confidence limit for one interesting parameter, with the $\chi^2$ fitting statistics. \label{tab:Swiftobs}}
\tablehead{
\colhead{ObsID}  & \colhead{Observed date}  & \colhead{Exposure time}  & \colhead{$N_{\rm H}$} & \colhead{$kT_{\rm in}$} & \colhead{$N_{\rm disk}$} & \colhead{$\Gamma$} & \colhead{State} \\
 \colhead{}  & \colhead{(day)} & \colhead{(s)}  &\colhead{($\rm 10^{22}\ cm^{-2}$)} & \colhead{(keV)}
}
\startdata
00886266000 & 2019-01-28 13:36:07 & 1838 & $0.74^{+0.03}_{-0.03}$    & $0.81^{+0.05}_{-0.05}$    & $650^{+170}_{-170}$          & $1.47^{+0.09}_{-0.10}$    & HS \\  
00886496000 & 2019-01-29 10:03:43 & 317  & $0.84^{+0.07}_{-0.06}$    & $0.79^{+0.09}_{-0.10}$    & $1400^{+600}_{-400}$         & $1.2^{+0.3}_{-0.2}$ 	 & HS \\  
00011107001 & 2019-01-30 02:34:04 & 834  & $0.78^{+0.04}_{-0.04}$    & $0.74^{+0.05}_{-0.05}$    & $2000^{+500}_{-400}$         & $1.35^{+0.12}_{-0.13}$ 	 & HS \\  
00088843001 & 2019-02-01 17:48:35 & 2033 & $0.76^{+0.02}_{-0.02}$    & $0.63^{+0.03}_{-0.03}$    & $6100^{+1200}_{-900}$        & $1.24^{+0.07}_{-0.07}$ 	 & HS \\  
00011107002 & 2019-02-03 16:06:14 & 896  & $0.85^{+0.04}_{-0.04}$    & $0.54^{+0.05}_{-0.06}$    & $8000^{+4000}_{-3000}$       & $1.72^{+0.09}_{-0.09}$ 	 & HS\\   
00011107003 & 2019-02-05 07:58:04 & 1365 & $0.91^{+0.04}_{-0.04}$    & $0.68^{+0.02}_{-0.02}$    & $16000^{+2000}_{-2000}$      & $1.84^{+0.14}_{-0.15}$ 	 & HIMS \\
00011107004 & 2019-02-07 01:36:28 & 1748 & $1.09^{+0.03}_{-0.03}$    & $0.742^{+0.014}_{-0.014}$ & $28900^{+2000}_{-1800}$      & $1.49^{+0.18}_{-0.19}$ 	 & HIMS \\
00011107005 & 2019-02-09 00:05:45 & 294  & $1.41^{+0.08}_{-0.08}$    & $0.76^{+0.04}_{-0.04}$    & $32000^{+8000}_{-6000}$ & $2.5^{a}$   & SIMS \\
\enddata
\end{deluxetable}
\footnotesize{\textbf{Note.} -- $\Gamma$ is the photon index of the {\tt powerlaw} model.} \newline
\footnotesize{$^a$  The photon index ($\Gamma$) is assumed
to be constant at the canonical value.}

\begin{deluxetable}{lccccccccc}
\tablecolumns{9}
\tablewidth{0pc}
\linespread{1.6}
\tablecaption{Best-fitting parameters for the \textit{Insight}-HXMT spectra in the hard and intermediate states, modelled with [{\tt diskbb+nthcomp+nthcomp}].\\  Error ranges represent the 90\% confidence limit for one interesting parameter, with the $\chi^2$ fitting statistics.  \label{tab:2nth}}
\tablehead{
\colhead{ExposureID} & \colhead{$N_{\rm H}$} & \colhead{$kT_{\rm in}$} & \colhead{$N_{\rm disk}$} & \colhead{$\Gamma_{1}$} & \colhead{$kT_{\rm e1}$} & \colhead{$\Gamma_{2}$} & \colhead{$kT_{\rm e2}$} & \colhead{$\chi^{2}$/dof} \\
 \colhead{}  &\colhead{($\rm 10^{22}\ cm^{-2}$)} & \colhead{(keV)} & \colhead{} & \colhead{} & \colhead{(keV)}  &\colhead{} & \colhead{(keV)} & \colhead{}
}
\startdata
P021400200201 & $0.73^{+0.02}_{-0.03}$   & $0.63^{+0.06}_{-0.07}$       & $560^{+310}_{-180}$       & $1.646^{+0.012}_{-0.024}$  & $55^{+5}_{-4}$       & $1.27^{+0.16}_{-0.15}$  & $10.7^{+1.3}_{-1.0}$   & $1238.3/1202$  \\
P021400200301 & $0.80^{+0.02}_{-0.02}$   & $0.53^{+0.03}_{-0.04}$       & $2200^{+900}_{-600}$      & $1.672^{+0.009}_{-0.013}$  & $44.6^{+1.8}_{-1.6}$ & $1.28^{+0.10}_{-0.10}$  & $9.8^{+0.7}_{-0.6}$    & $1328.5/1202$  \\
P021400200401 & $0.76^{+0.02}_{-0.02}$   & $0.59^{+0.03}_{-0.03}$       & $1800^{+400}_{-300}$      & $1.701^{+0.011}_{-0.022}$  & $44.0^{+1.9}_{-1.7}$ & $1.39^{+0.10}_{-0.08}$  & $10.1^{+0.8}_{-0.6}$   & $1430.7/1202$  \\
P021400200402 & $0.74^{+0.03}_{-0.03}$   & $0.61^{+0.05}_{-0.05}$       & $1800^{+600}_{-400}$      & $1.69^{+0.02}_{-0.40}$     & $45^{+16}_{-7}$      & $1.43^{+0.29}_{-0.17}$  & $10.4^{+2.6}_{-1.6}$   & $1185.3/1202$  \\
P021400200403 & $0.75^{+0.02}_{-0.02}$   & $0.57^{+0.04}_{-0.05}$       & $1900^{+800}_{-500}$      & $1.723^{+0.012}_{-0.018}$  & $48^{+3}_{-2}$       & $1.37^{+0.10}_{-0.10}$  & $10.5^{+0.8}_{-0.7}$   & $1297.4/1202$  \\
P021400200404 & $0.76^{+0.03}_{-0.02}$   & $0.58^{+0.05}_{-0.05}$       & $1900^{+900}_{-500}$      & $1.722^{+0.014}_{-0.019}$  & $44.1^{+2.0}_{-1.7}$ & $1.37^{+0.11}_{-0.13}$  & $10.0^{+0.8}_{-0.8}$   & $1307.6/1202$  \\
P021400200405 & $0.79^{+0.04}_{-0.03}$     & $0.52^{+0.04}_{-0.06}$       & $3500^{+2500}_{-1300}$    & $1.727^{+0.015}_{-0.026}$  & $46^{+3}_{-2}$       & $1.40^{+0.12}_{-0.12}$  & $10.5^{+0.8}_{-0.8}$   & $1358.6/1202$  \\
P021400200407 & $0.76^{+0.04}_{-0.03}$     & $0.59^{+0.05}_{-0.06}$       & $2400^{+1300}_{-800}$     & $1.70^{+0.04}_{-0.40}$     & $47^{+7}_{-13}$      & $1.55^{+0.24}_{-0.15}$  & $11.9^{+1.8}_{-1.6}$   & $1212.3/1202$  \\
\enddata
\end{deluxetable}
\footnotesize{\textbf{Note.--}
$^a$ The temperature went below the lower limit and we fixed it at that limit.}

\begin{deluxetable}{lccccccccc}
\tablecolumns{9}
\tablewidth{0pc}
\linespread{1.6}
\tablecaption{Best-fitting parameters for the \textit{Insight}-HXMT spectra in the hard and intermediate states, modelled with [{\tt simplcut*diskbb+nthcomp}].\\  Error ranges represent the 90\% confidence limit for one interesting parameter, with the $\chi^2$ fitting statistics.  \label{tab:simplcut}}
\tablehead{
\colhead{ExposureID} &  \colhead{$N_{\rm H}$} & \colhead{$kT_{\rm in}$} & \colhead{$N_{\rm disk}$}  & \colhead{$\Gamma_{1}$}  & \colhead{$kT_{\rm e1}$} & \colhead{frac} & \colhead{$\Gamma{_2}$} & \colhead{$kT_{\rm e2}$} & \colhead{$\chi^{2}$/dof} \\
 \colhead{} &\colhead{($\rm 10^{22}\ cm^{-2}$)} & \colhead{(keV)} & \colhead{} & \colhead{} & \colhead{(keV)} &\colhead{} &\colhead{} & \colhead{(keV)} & \colhead{}
}
\startdata
P021400200201 &  $0.717^{+0.018}_{-0.017}$  &   $0.65^{+0.06}_{-0.04}$        &  $650^{+220}_{-170}$        &  $< 1.2$              &  $24.8^{+12.3}_{-0.8}$ &  $0.089^{+0.220}_{-0.011}$  & $1.614^{+0.018}_{-0.005}$     &  $15.6^{+1.0}_{-0.7}$    & $1247.0/1202$ \\
P021400200301 &  $0.775^{+0.016}_{-0.016}$  &   $0.56^{+0.02}_{-0.02}$        &  $2200^{+500}_{-400}$       &  $< 1.1$            &  $22.6^{+0.8}_{-0.5}$  &  $0.053^{+0.005}_{-0.005}$  & $1.632^{+0.003}_{-0.003}$     &  $14.3^{+0.5}_{-0.4}$    & $1369.6/1202$ \\
P021400200401 &  $0.745^{+0.013}_{-0.011}$  &   $0.60^{+0.02}_{-0.02}$        &  $2000^{+400}_{-300}$       &  $< 1.1$              &  $22.1^{+1.0}_{-0.4}$  &  $0.049^{+0.013}_{-0.004}$  & $1.6574^{+0.0026}_{-0.0014}$  &  $13.9^{+0.4}_{-0.6}$    & $1483.2/1202$ \\
P021400200402 &  $0.729^{+0.017}_{-0.010}$  &   $0.61^{+0.03}_{-0.03}$        &  $2000^{+500}_{-300}$       &  $< 1.4$              &  $22^{+18}_{-2}$       &  $0.050^{+0.331}_{-0.003}$  & $1.651^{+0.050}_{-0.006}$     &  $13.3^{+1.8}_{-1.2}$    & $1202.4/1202$ \\
P021400200403 &  $0.729^{+0.016}_{-0.015}$  &   $0.60^{+0.03}_{-0.03}$        &  $2000^{+500}_{-400}$       &  $< 1.3$              &  $22.8^{+8.2}_{-0.5}$  &  $0.045^{+0.164}_{-0.005}$  & $1.664^{+0.038}_{-0.003}$     &  $14.5^{+0.5}_{-0.4}$    & $1323.7/1202$ \\
P021400200404 &  $0.736^{+0.002}_{-0.002}$  &   $0.60^{+0.03}_{-0.03}$        &  $2100^{+500}_{-600}$       &  $< 1.4$              &  $22.0^{+11}_{-0.5}$   &  $0.046^{+0.259}_{-0.005}$  & $1.672^{+0.047}_{-0.003}$     &  $14.1^{+0.6}_{-0.4}$    & $1336.3/1202$ \\
P021400200405 &  $0.77^{+0.04}_{-0.03}$     &   $0.55^{+0.04}_{-0.04}$        &  $3400^{+1500}_{-1000}$     &  $< 1.1$             &  $22.6^{+3.0}_{-0.6}$  &  $0.035^{+0.007}_{-0.006}$  & $1.672^{+0.003}_{-0.003}$     &  $14.6^{+0.5}_{-0.5}$    & $1369.6/1202$ \\
\enddata
\end{deluxetable}
\footnotesize{\textbf{Note.--}
$\Gamma_{1}$ is the photon power law index of {\tt simplcut}; $kT_{\rm e1}$ is the electron temperature of {\tt simplcut} in units of keV; frac is the scattered fraction; $\Gamma_{2}$ is the asymptotic power-law photon index of {\tt nthcomp}; $kT_{\rm e2}$ is the electron temperature of {\tt nthcomp} in units of keV.} \\
\footnotesize{$^a$ The temperature went below the lower limit and we fixed it at that limit.}
\end{document}